\documentclass[11pt,oneside,letterpaper]{article}
\usepackage{amssymb}
\usepackage{amsmath}
\usepackage[dvips]{graphicx}
\usepackage{setspace}
\usepackage{amsfonts}
\usepackage{fancyhdr}
\usepackage{xcolor}
\usepackage{graphicx}
\usepackage{rotating}
\usepackage{comment}
\usepackage{color}
\usepackage{cite}
\usepackage{braket}
\usepackage{bbm}
\usepackage{float}
\definecolor{dolphingreen}{rgb}{0,0.74,0.63}
\definecolor{dolphinorange}{rgb}{0.98,0.53,0.098}
\definecolor{purple}{rgb}{0.4,.2,0.7}
\newcommand{\p}{\partial}

\newcommand{\f}{\frac}

\usepackage[colorlinks=true,citecolor=dolphinorange,linkcolor=black,urlcolor=dolphingreen]{hyperref}
\usepackage{pdfsync}

\makeatletter
\newcommand*{\defeq}{\mathrel{\rlap{%
                     \raisebox{0.3ex}{$\m@th\cdot$}}%
                     \raisebox{-0.3ex}{$\m@th\cdot$}}%
                     =}
\makeatother

\DeclareMathOperator{\e}{\epsilon}
\DeclareMathOperator{\Tr}{Tr}
\def\be{\begin{eqnarray}}
\def\ee{\end{eqnarray}}

\newcommand{\<}{\langle}

\newcommand{\bea}{\begin{eqnarray}}
\newcommand{\eea}{\end{eqnarray}}
\def\ben{\begin{equation}}
\def\een{\end{equation}}
\def\bne{\begin{equation}}
\def\ene{\end{equation}}

\let\a=\alpha \let\b=\beta \let\g=\gamma \let\d=\delta \let\e=\varepsilon
\let\z=\zeta   \let\k=\kappa
\let\l=\lambda \let\m=\mu \let\n=\nu  \let\p=\phi \let\r=v
 \let\t=\tau

\let\w=\omega  \let\D=\Delta  \let\L=\Lambda

\let\f=\frac

\def\ba{\begin{array}}
\def\ea{\end{array}}
\def\del{\partial}

\allowdisplaybreaks  

\renewcommand*{\defeq}{\mathrel{\vcenter{\baselineskip0.5ex \lineskiplimit0pt
                     \hbox{\scriptsize.}\hbox{\scriptsize.}}}%
                     =}

\newcommand{\textoverline}[1]{$\overline{\mbox{#1}}$}

\numberwithin{equation}{section}

\begin{document}
\onehalfspacing

\begin{center}

~
\vskip5mm

{\LARGE  {\textsc{T\textoverline{T} in A}\MakeLowercase{d}\textsc{S$_2$ and Quantum Mechanics}}}

\vskip10mm
David J. Gross$^1$, Jorrit Kruthoff$^2$, Andrew Rolph$^{1,3}$, and Edgar Shaghoulian$^4$

\vskip5mm

{\it 1)  Kavli Institute for Theoretical Physics,\\
University of California, Santa Barbara, CA 93106}\vskip3mm
{\it 2)  Institute for Theoretical Physics Amsterdam and $\D$ Institute for Theoretical Physics, University of Amsterdam, Science Park 904, 1098 XH Amsterdam, The Netherlands} \vskip3mm
{\it 3) Martin A. Fisher School of Physics,\\
Brandeis University, Waltham, MA 02453, USA}
\vskip3mm
{\it 4) Department of Physics, Cornell University, Ithaca, New York, USA }

\vskip5mm


\end{center}

\vspace{4mm}

\begin{abstract}
\noindent Quantum field theory in zero spatial dimensions has a rich infrared landscape and a universal ultraviolet, inverting the usual Wilsonian paradigm. For holographic theories this implies a rich landscape of asymptotically AdS$_2$ geometries. Isolating the interiors of these spacetimes suggests a study of the analog of the $T\overline{T}$ deformation in quantum mechanics, which we pursue here.  An equivalent description of this deformation in terms of coupling to worldline gravity is proposed, and applications to quantum mechanical systems -- including the Schwarzian theory -- are studied.
 \end{abstract}

\pagebreak
\pagestyle{plain}

\setcounter{tocdepth}{2}
{}
\vfill
\tableofcontents

\section{Introduction}
The techniques of the ordinary Wilsonian renormalization group can be applied to theories of quantum mechanics, i.e. $(0+1)$-dimensional quantum field theories, but they result in a very dramatic feature. The universality of infrared physics, wherein lies the power of low-energy effective field theory, is totally lost. Rather than having a small handful of relevant operators that can be written down to deform free field Lagrangians, every (non-derivative, nonsingular) operator appears to be relevant. This allows the possibility of a rich infrared landscape not available in higher-dimensional theories. On the other hand, there is a universality of the ultraviolet of quantum mechanics due to the paucity of irrelevant operators. This picture suggests exploring the diversity of the infrared landscape and the robustness of the ultraviolet landscape.

There is a wide class of integrable deformations of quantum mechanics parameterized by arbitrary functions. These deformations can be considered as transformations of the Hamiltonian $H \rightarrow F(H)$. They can also be written as first-order flows $\partial H/\partial\lambda = f(H)$ or $\partial L_E/\partial\lambda = f(T)$, where $L_E$ is a Euclidean Lagrangian and $T$ the stress tensor, which is really just a scalar in $d=1$. The operators providing the deformation are manifestly well-defined ``composite" operators. The flows are integrable in the sense that the deformed eigenvalues of the Hamiltonian are immediately calculable in terms of the original eigenvalues, and the eigenfunctions are unchanged. An analysis of these more general deformations and their application to interesting quantum mechanical models will be considered in future work \cite{paperfuture}. 

In this paper, we will focus on a particular $f(H)$ deformation important for holography. In the context of AdS/CFT, the Wilsonian picture implies a rich landscape of asymptotically AdS$_2$ geometries. One example of such a geometry is the Anninos-Hofman centaur, which embeds the static patch of two-dimensional de Sitter spacetime into an asymptotically AdS$_2$ spacetime \cite{Anninos:2017hhn, Anninos:2018svg}. Isolating and studying these infrared geometries by excising the ultraviolet region would provide a potential route to holography for general spacetimes, such as de Sitter space or other accelerated cosmologies, through the rigorous starting point of AdS/CFT. Furthermore, through the examples of $D0$-brane quantum mechanics and BFSS theory, it is clear that quantum mechanics -- without the complication of spatial locality and Lorentz invariance that comes from quantum field theory -- already contains \emph{local} emergent spacetime with black holes. Many of the puzzles of quantum gravity are therefore directly accessible in theories of quantum mechanics.

The strategy we pursue here is to derive and study the analog of the $T\overline{T}$ deformation \cite{Zamolodchikov:2004ce, Smirnov:2016lqw, Cavaglia:2016oda} for one-dimensional quantum mechanical theories. This is an irrelevant deformation that is proposed to be dual to a sharp radial cutoff in anti-de Sitter space \cite{McGough:2016lol}, which performs the excision required to isolate the infrared geometry. In addition to allowing us to access a diverse landscape of geometries, the $T\overline{T}$ deformation in $d=1$ is a universally well-defined composite operator available in any quantum-mechanical theory. Furthermore, it will explicitly preserve all symmetries of the original theory (including supersymmetry) and will provide a universal way to couple quantum mechanics to worldline gravity. 

\subsection*{\it Summary}

In section \ref{ttbar} we derive the deforming operator corresponding to a finite Dirichlet cutoff in AdS$_2$. We begin in section \ref{rewrite} with Jackiw-Teitelboim (JT) gravity in AdS$_2$ \cite{JACKIW1985343, TEITELBOIM198341}, obtaining the operator via dimensional reduction of the $T\overline{T}$ operator in AdS$_3$. In section \ref{general2d} we derive the deforming operator for a general dilaton-gravity theory with matter fields directly in AdS$_2$ by applying the algorithm of \cite{Hartman:2018tkw}. Such theories can capture exotic geometries like the Anninos-Hofman centaur, which contains an infrared de Sitter region with its cosmological horizon accessible to probes from the ultraviolet AdS$_2$ boundary. By applying the flow derived in section \ref{general2d} the de Sitter region can be isolated. We furthermore show that the flow in the de Sitter region is a $d=1$, static patch version of the $T\overline{T} + \Lambda_2$ flow of \cite{Gorbenko:2018oov}. In section~\ref{shocksbulk} we investigate chaos in JT gravity at finite cutoff, calculating the Lyapunov exponent and showing it to be unaffected. 

In section \ref{qmsec} we study properties of the flow triggered by the deforming operator of section \ref{ttbar} directly in quantum-mechanical theories. In section \ref{symms} we show that all symmetries, including supersymmetry, are preserved along the flow. In section \ref{uvqm} we solve for the deformed ultraviolet Lagrangians obtained by flowing generic quantum-mechanical theories. In all cases we find that the ultraviolet theory can be written -- in a particular gauge -- as a worldline action with a cosmological constant or background electric field. 

In section \ref{schwarziansec}, we apply our deformation to the Schwarzian theory. We derive the action of the deformed theory in section~\ref{sec:DSA}. To compare to the bulk, we recast the finite cutoff dictionary in terms of a gauge-invariant fixed-length boundary condition as is commonly used in the Schwarzian literature. Section \ref{chaosbdry} is devoted to a computation of the Lyapunov exponent in the deformed Schwarzian theory. In section \ref{thermosch} we compute the deformed density of states and give an exact expression for the deformed partition function.  

In section \ref{nonpert} we provide a nonperturbative definition of our deformation in terms of coupling to a theory of worldline gravity. This can be understood as the one-dimensional version of the proposal that a $T\overline{T}$ deformation of a two-dimensional QFT can be nonperturbatively defined as coupling the theory to Jackiw-Teitelboim gravity with vanishing cosmological constant \cite{Dubovsky:2018bmo}.

Appendix \ref{tangent} looks at distinct quantum-mechanical flows triggered by the operator $T^2$. The flow of a free particle is shown to result in a deformed Lagrangian that is functionally identical to that obtained by a $T\overline{T}$ deformation of 2d Yang-Mills. Appendix \ref{ccharges} shows that the deformation of two-dimensional interacting scalars by the operator proposed in section \ref{rewrite} leads to a Nambu-Goto action in a nontrivial target space metric.

A few clarifying comments about conventions are in order. We will often switch between the Hamiltonian $H$ and the stress tensor $T$, which is just a scalar for $d=1$. The only distinction meant here is that $H$ is an operator written in terms of fields and their conjugate momenta, while $T$ is in terms of fields and their derivatives. We will mostly work in Euclidean signature, writing $S_E$ and $L_E$ for Euclidean actions and Lagrangians. 


\section{$T\overline{T}$ flow in AdS$_2$}\label{ttbar}
In this section we will derive the particular deformation that corresponds to a finite Dirichlet cutoff in AdS$_2$. For $d>1$, the deformation that corresponds to finite cutoff for Einstein gravity in AdS$_{d+1}$ is always made up of a quadratic combination of the stress tensor. The natural guess for $d=1$ is then to write a flow that uses the only possible quadratic combination available: $\partial  S_E/\partial\lambda = \int d\tau \,T^2$. For some analysis of this deformation, see Appendix \ref{tangent}. The conclusion is that this flow does not seem to capture the physics of simple gravitational theories at finite cutoff that we are accustomed to from the higher-dimensional examples: the differential equation for the energy levels does not take the form of Burger's equation, and the solution does not have a square-root singularity.  

To make a sharp statement we need to specify a bulk theory in AdS$_2$ for which we want to extract the deforming operator that gives the physics at finite Dirichlet cutoff. For a general bulk theory we will have to use the algorithm from  \cite{Hartman:2018tkw} to derive the deforming operator in the dual quantum mechanics (we will do this in section \ref{general2d}). For Jackiw-Teitelboim gravity, we can alternatively exploit the relation to three-dimensional gravity in the spherically symmetric sector to obtain the deformation through dimensional reduction \cite{Castro:2008ms}. We now turn to this. 

We begin with three-dimensional Einstein gravity with negative cosmological constant and the usual boundary terms
\be\label{3dgrav}
S_E = -\f{1}{16\pi G}\int d^3x \sqrt{g} \left(R+2\right)-\f{1}{8\pi G}\int d^2 x  \sqrt{g^{0}} \left(K-1\right)
\ee
and parameterize the metric as 
\be
ds^2 = g^{(2)}_{\a\b}(x^\a)dx^\a dx^\b+\Phi^2(x^\a) d\phi^2=  g_{rr}dr^2 + r^2 \g_{\tau \tau}d\tau^2+\Phi^2(r,\t) d\phi^2 \,,
\ee
with $\phi \sim \phi+1$. In particular we have assumed a $U(1)$ and  $\mathbb{Z}_2$ reflection symmetry for $\phi$. Asymptotically AdS$_3$ metrics require $\Phi \rightarrow \mathcal{O}(r)$ at large $r$. Solutions that fit into this ansatz include spinless BTZ black holes:
\be
ds^2 = (r^2-r_+^2)d\t^2 + \f{dr^2}{r^2-r_+^2}+r^2  d\phi^2\,,\quad M = \f{r_+^2 }{16\pi G}\,,\quad \widetilde{E} = \f{r_c }{8\pi G}\left(1-\sqrt{1-\f{r_+^2}{r_c^2}}\right),
\ee
where we have written the usual mass $M$ of the solutions, and the energy $\widetilde{E}$ at finite cutoff $r_c$. The energy is computed as $\widetilde{E}=\int d\p \sqrt{g_{\p\p}} \,\widetilde{T}_{\a\b}u^\a u^\b$, with $u^{\a}$ a normal vector to the radial cutoff surface. 

To dimensionally reduce the action above along the $\phi$ circle, we use the following relation between the three and two dimensional Ricci scalar and extrinsic curvature,
\be\label{3dto2d}
R_{3d} = R_{2d} - 2 \Phi^{-1}\square \Phi,\quad K_{2d} = K_{1d} + n^{\mu}\partial_{\mu} \log \Phi,
\ee
with $n^{\mu}$ a normal vector to the one-dimensional boundary. Upon plugging this in \eqref{3dgrav}, the second terms in both equations in \eqref{3dto2d} cancel and we end up with the 2d action,
\be
S_E = -\f{1}{16\pi G}\int d^2x \sqrt{g^{(2)}}\,\Phi \left(R+2\right)-\f{1}{8\pi G}\int \sqrt{r^2 \g_{\tau\tau}}\,\Phi \left(K-1\right),
\ee
where all geometric quantities are now in one lower dimension. The effective two-dimensional Newton constant is the dimensionless ratio $\Phi/G$. This is JT gravity with an appropriate set of boundary terms. Thus all solutions of 3d gravity with spherical and $\mathbb{Z}_2$ symmetry lead to solutions of JT gravity. For example the BTZ black holes lead to the two-dimensional black holes of JT gravity:
\be
ds^2 = (r^2-r_+^2)d\t^2 + \f{dr^2}{r^2-r_+^2}\,,\quad \Phi = r \,,
\ee
and all solutions of JT gravity uplift to solutions of 3d gravity. 

\subsection{Dimensional reduction of $T \overline{T}$ in $d=2$ }\label{rewrite}
Before we dimensionally reduce the $T\overline{T}$ flow of CFT$_2$, we will first write it in a different-looking but equivalent form. (An infinite number of such rewritings are possible.) For a flow of a CFT given by
\be\label{ttbarflow}
\f{\partial S_E}{\partial \l} = 8\int d^2 x \sqrt{\g}\,T\overline{T},
\ee
we have $T^\m_\m = -16\l T\overline{T}=-2\l(T_{ij}T^{ij}-(T^i_i)^2)$ along the flow. We solve this equation for $T^\p_\p$ to get
\be
T^\p_\p = \f{T^\t_\t+4\l T_{\t\p}T^{\t\p}}{4\l T^\t_\t - 1}\,.
\ee
Plugging this into the flow equation \eqref{ttbarflow} gives 
\be\label{flowrewrite}
\f{\partial S_E}{\partial \l} = \int d^2 x \sqrt{\g}\,\left(\f{(T^\t_\t)^2+T_{\t\p}T^{\t\p}}{1/2-2\l T^\t_\t}\right).
\ee
A few comments are in order about this equation. The first is that it involves composite operators of arbitrary powers. For the holographic application we are interested in, these are well-defined due to large-$c$ factorization. At finite $c$, this may be well-defined by its relation to the $T\overline{T}$ operator. In particular, as long as $T^\m_\m \propto T\overline{T}$ as operators along this flow, then the operator above is equal to $T\overline{T}$ and therefore well-defined. In particular, to agree with the results of $T\overline{T}$ deformation it will also have to obey factorization as we will see below. The operator can be made manifestly well-defined by changing it slightly. In particular, we can rewrite $T^\t_\t \rightarrow \int T^\t_\t=E$ and $T_{\t\p}=T^{\t\p} \rightarrow \int T^{\t\p}= iJ$, where we set the circle size to one. Then we will have composite operators built out of powers of energy and momentum, which are manifestly well-defined and factorize in eigenstates of energy and momentum. The deformation written this way is not manifestly local. In appendix \ref{ccharges} we discuss deforming non-conformal theories by such an operator, which is \emph{not} the same as $T\overline{T}$.

The second comment concerns the presence of a pole in the deforming operator due to the $1/2 - 2\l T^\t_\t$ denominator. This is only an issue in trying to understand the flow nonperturbatively. A priori, one can define this in various ways, including analytic continuation in $\l$. However, this pole is signaling the complexification of the energy levels. It occurs for states with energy $T^\t_\t(\l) = 1/4\l$. Jumping ahead to \eqref{energies}, we see that this happens when the square root vanishes. By slightly increasing or decreasing $\l$ we can make the energy of this state become complex, which regularizes the singularity. The only states for which this is not true are the double zeros of the square root, which are extremal states with $E_0 = J = 1/4\l$. Such states are protected and do not flow;
the pole is cancelled by the zero in $E^2 -  J^2$ in the numerator. 

Using $\langle T^\t_\t\rangle = E$ and $\langle T_{\t\p}\rangle=\langle T^{\t\p}\rangle=iJ$ in  energy-momentum eigenstates and assuming factorization, we obtain the following equation for the energy levels:
\be
\f{\partial E}{\partial \l} = \f{E^2 - J^2}{1/2 - 2 \l E}\,,
\ee
which has the usual solution
\be
E(\l) = \f{1}{4\l}\left(1-\sqrt{1-8 \l E_0 + 16 \l^2 J^2}\right).\label{energies}
\ee
Since $T^\p_\p$ has been eliminated in the right-hand-side of the flow equation \eqref{flowrewrite}, it will now be easier to dimensionally reduce. We set $T_{\t\p} = T^{\t\p} = 0$ and call $T^\t_\t \defeq T$ to get 
\be\label{floweq}
\f{\partial S_E}{\partial \l} = \int d\t  \,\, \f{ T^2}{1/2-2\l T}\,,
\ee
where we have fixed to a metric of unit lapse. This leads to a differential equation for the energy levels:
\be\label{enfloweq}
\f{\partial E}{\partial \l} = \f{E^2}{1/2-2\l E} \implies E(\l) = \f{1-\sqrt{1-8E_0\l}}{4\l}\,.
\ee
This agrees precisely with the energy of the two-dimensional black holes at finite cutoff upon the identification  $\l = 2\pi G/r_c^2$. (Recall that to translate the bulk answer to a boundary quantity we need to rescale by an additional factor of $r_c$, see equation \eqref{dict}.) If we modified the source for the dilaton as $\Phi = r \Phi_r$, which would correspond to circle size $\Phi_r$ in the three-dimensional picture, the flow equations \eqref{floweq} - \eqref{enfloweq} would remain true with the modified identification $\l = 2\pi G/(\Phi_r r_c^2)$.

The energies $E_0$ are the energies of the initial theory, which are obtained from $S_E$ in the limit $\l \to 0$. This theory should be thought of as a nearly CFT$_1$ which is dual to nearly AdS$_2$. An example of such a theory is the Schwarzian theory, which we will discuss in detail in section \ref{schwarziansec}.

Setting $E_0=1$ and rescaling $\l \rightarrow \l/2$ shows that $E(\l)$ is the generating function of the Catalan numbers. In other words, the deformed energy levels (and as we will shortly see, the deformed Hamiltonian) solve the equation
\be
E(\l) = E_0 + 2\l E(\l)^2 \implies E(\l) = \sum_{n=0}^\infty C_n (2\l)^n E_0^{n+1}, \qquad C_n = \f{(2n)!}{(n+1)!n!}\,.
\ee

\subsection{General dilaton-gravity with matter}\label{general2d}
The deformation found by dimensional reduction to $d=1$ in the previous subsection can also be found directly in $d=1$ using the technique in \cite{Hartman:2018tkw}. To allow for more general bulk solutions, we will turn on a potential $U(\Phi)$ for the dilaton $\Phi$. This accommodates theories like the CGHS model \cite{Callan:1992rs} and admits an embedding of the de Sitter cosmic horizon within an asymptotically AdS$_2$ spacetime \cite{Anninos:2017hhn, Anninos:2018svg}. At the end of this section we will also add  matter. The action is
\be
S_E = -\frac{1}{2\k^2} \int_M d^2 x \sqrt{g} \left(\Phi R -2U(\Phi)\right) - \frac{1}{\k^2}\int_{\partial M} d\t \sqrt{g^0}\Phi \left(K-1\right),
\ee
with $g^0 = g_{\t\t}$ the induced metric on the boundary and $K$ its extrinsic curvature. To have an asymptotic AdS$_2$ geometry at large negative values for the dilaton would mean $U(\Phi) \to -\Phi$ as $\Phi \to -\infty$. The boundary term $\k^{-2}\int d\t \sqrt{g^0}\, \Phi$ ensures we obtain finite quantities in such a situation, although the analysis below is more general. The equations of motion for the dilaton gives $R=2U'$. The equations of motion of the metric are 
\be
\left(g_{\mu\nu} \nabla^2 - \nabla_{\mu}\nabla_{\nu}\right)\Phi + g_{\mu\nu}U(\Phi)= 0\,.
\ee
We consider a gauge in which the metric is diagonal and focus on static dilatons:
\be\label{sol2d}
\Phi = \Phi(r),\quad  ds^2 = N^2(r) d\t^2 + \frac{dr^2}{N^2(r)}.
\ee
A simple solution to the equations of motion of JT gravity is given by \cite{GrumillerMcNees}
\be
\Phi = r\Phi_r, \quad N^2 = f(\Phi)\left(1 - \frac{2M\k^2 }{\Phi_r f(\Phi)}\right)
\ee
with 
\be\label{deff}
f(\Phi) = -\f{2}{\Phi_r^2}\int^{\Phi} U(x)\;dx\,.
\ee
This solution describes a two dimensional black hole whenever $f(\Phi) > 0$ and has horizons at $f(\Phi) = 2M\k^2 /\Phi_r $. The constant $M$ plays the role of the mass of the solution. Notice that these solutions also possess a Killing vector $\partial_{\t}$. The renormalized Brown-York stress scalar and canonical momentum of the dilaton are given by 
\begin{align}
\widetilde{T}^{\t\t} &= \frac{2}{\sqrt{g^0}}\f{\d S_E}{\d g^0}=\frac{2}{\sqrt{g^0}}\left(\pi^{\t\t} + \frac{\sqrt{g^0}}{2}\f{1}{g^0}\frac{\Phi}{\k^2}\right),\\
\widetilde{O} &= \frac{1}{\sqrt{g^0}}\f{\d S_E}{\d \Phi}=\frac{1}{\sqrt{g^0}}\left(\pi_{\Phi} + \frac{\sqrt{g^0}}{\k^2}\right).
\end{align}
which are obtained by varying the on-shell action with respect to $\g_{\t\t}$ and $\Phi$. The quasilocal energy of our solution \eqref{sol2d} at finite radial cutoff $r = r_c$ is
\be\label{energyAdS2}
E_{\rm bulk} = \widetilde{T}^{\t}_{\t} =\frac{\Phi_r r_c}{\k^2}\left(1 - \frac{1}{r_c} \sqrt{f(\Phi_0)-2M\k^2 /\Phi_r}\right),
\ee
where $\Phi_0 = \Phi(r = r_c)$. To derive the deformation of the one-dimensional boundary theory that reproduces these energy levels, we proceed as in higher dimensions by analyzing the Hamiltonian constraint. For our theory it is given as
\be \label{hamcon}
\mathcal{H} =  \k^2 \widetilde{O} \widetilde{T}^{\tau}_{\tau} - \Phi_0\widetilde{O} - \widetilde{T}^{\t}_{\t} + \frac{1}{\k^2 }\left(\Phi_0 + U(\Phi_0)\right) = 0\,. 
\ee
For a purely AdS$_2$ potential $U(\Phi_0) = -\Phi_0$, the last two terms in the Hamiltonian constraint cancel. The deformation can be derived from this equation as discussed in \cite{Hartman:2018tkw}, where in this case we will also flow the source for the dilaton. The bulk flow equation is given as 
\be
\f{d}{dr_c} W[g^0(r_c), \Phi_0(r_c)] = \int d\t \sqrt{g^0}\left(\f{1}{2} \widetilde{T}^{\t\t}\partial_{r_c} g^0(r_c)+\widetilde{O}\,\partial_{r_c} \Phi_0\right).
\ee
The dictionary to translate between bulk variables and EFT variables is given as
\be\label{dict}
g_{\t\t}^0 = r_c ^2\g_{\t\t},\quad \widetilde{T}_{\t\t} = r_c T_{\t\t},\quad
 \Phi_0 =  r_c\Phi_r, \quad \widetilde{O} =  r_c^{-2} O\,.
\ee
The choice for the sources is made by demanding finiteness as $r_c\rightarrow \infty$, although at finite $r_c$ other choices are possible. 
Following the steps in \cite{Hartman:2018tkw} leads to
\be
\f{\partial}{\partial r_c} S_{EFT} = \int d\tau \sqrt{\g}\,\Theta
\ee
with $\Theta$ given by $\Theta = \widetilde{T}^{\t}_{\t}+\Phi_0\widetilde{O}$. We use the Hamiltonian constraint \eqref{hamcon} to rewrite this as 
\bea
\Theta = \k^2  \widetilde{O}\widetilde{T}^{\t}_{\t} + \frac{1}{\k^2}\left(\Phi_0 + U(\Phi_0)\right),
\eea
which -- using the dictionary to translate to field theory variables -- is given as
\be
\Theta= \frac{\k^2}{r_c^3}OT^{\tau}_{\t} + \frac{1}{\k^2}\left(r_c\Phi_r + U(r_c\Phi_r)\right).
\ee
The flow of the effective action is thus
\be
r_c \frac{\partial S_{EFT}}{\partial r_c} = \int d\t \sqrt{\g} \left( \frac{\k^2}{r_c^2} OT^{\t}_{\t} + \frac{r_c^2}{\k^2}\left(\Phi_r+ r_c^{-1} U(r_c\Phi_r)\right) \right).
\ee
Using the Hamiltonian constraint one more time in the form
\be
O = \f{r_c T^\t_\t - r_c^2\k^{-2}(r_c \Phi_r + U(r_c \Phi_r))}{\k^2 r_c^{-1} T^\t_\t-r_c \Phi_r}
\ee
to trade out $O$ from the flow equation, and using $\l = \k^2 /(4\Phi_r r_c^2)$, we can write our flow as
\be
r_c\frac{\partial S_{EFT}}{\partial r_c} = \int d\t \sqrt{\g}\,\left(\f{-16r_c T^2\l^2+r_c+  U(r_c \Phi_r)/\Phi_r}{4 \l r_c-16 r_cT\l^2}\right)
\ee
\be\label{KSflow}
\implies
\frac{\partial S_{EFT}}{\partial \l} =\int d\t \sqrt{\g}\left(\f{T^2-\left(1+(r_c \Phi_r)^{-1} U(r_c \Phi_r)\right) /(16\l^2)}{1/2-2\l T}\right),
\ee
where we again defined $T^\t_\t \defeq T$. Setting $U(x)=-x$ gives the flow we considered for JT gravity.

The flow equation for the effective action can be turned into a flow equation for the energy levels of the deformed quantum mechanics:
\be
\frac{\partial E}{\partial \l} = \frac{E^2 - \left(1 + \f{\sqrt{\a\l}}{\Phi_r}U(\f{\Phi_r}{\sqrt{\a\l}})\right)/(16\l^2)}{1/2 - 2\l E},
\ee
with $\a = 4 \Phi_r/\k^2$. The solutions to this flow equation are 
\be
E(\l) = \frac{1}{4\l}\left(1 - \sqrt{\a\l f(\Phi_r/\sqrt{\a \l}) - 8 \l E_0} \right),
\ee
where $f(x)$ is defined as in \eqref{deff}. The JT energy levels are obtained by setting $f(x) = x^2/\Phi_r^2$. 

\subsection*{\it Dissecting the Anninos-Hofman centaur}

An interesting application of this flow is to the AdS$_2$-dS$_2$ centaur geometry \cite{Anninos:2017hhn, Anninos:2018svg}. This solution arises for certain special potentials $U(\Phi)$ and is an embedding of the static patch of dS$_2$ inside of an asymptotically AdS$_2$ spacetime. The remarkable feature of this two-dimensional embedding is that the physics of the de Sitter horizon is accessible to the AdS boundary.  Using the flow \eqref{KSflow} we can then explicitly isolate the static patch region. In that region $f(x) \approx -x^2/\Phi_r^2$, and the resulting flow is precisely a two-dimensional, static patch version of the flow in \cite{Gorbenko:2018oov}. (The static patch case is also currently under consideration \cite{evatalk}.) In this case we do not need to consider the proposed $\Lambda_2$ flow, and instead turn on a source  that gradually changes the geometry from AdS$_2$ to dS$_2$.

\subsection*{\it Adding matter}
We can also consider the flow equation for gravitational theories in AdS$_2$ with additional matter. We will consider a matter sector that does not couple directly to the dilaton. (For two-dimensional dilaton-gravity models that descend from higher-dimensional theories with free matter, there will be a coupling to the dilaton due to the dimensional reduction over the transverse directions.) We also assume that we do not add any new boundary counterterms for the bulk matter. The Hamiltonian constraint of this section gets modified by the addition of the radial-radial component of the bulk matter stress tensor $\tilde{t}^r_r$. This feeds into the flow considered in this section by an additive term:
\be
\frac{\partial S}{\partial \l} =\int d\t \,\,\left(\f{T^2-\left(1+(r_c \Phi_r)^{-1} U(r_c \Phi_r)\right) /(16\l^2)}{1/2-2\l T} - \frac{ \tilde{t}^r_r r_c /(4\l)}{1/2 - 2\lambda T}\right).
\ee
 As in \cite{Hartman:2018tkw} the matter stress tensor needs to be processed into operators and sources of the quantum-mechanical theory. 
 The curious difference between $d=1$ and $d>1$ is the direct coupling between the bulk matter stress tensor and the Brown-York stress tensor. 
 
\subsection{Chaos in JT gravity at finite cutoff}\label{shocksbulk}
\begin{figure}[h!]
\begin{center}
\includegraphics[scale=0.1]{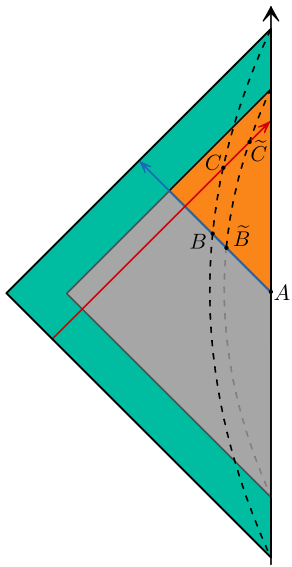}
\caption{Shockwaves in JT gravity from finite cutoff. The aqua green and orange regions correspond to black hole exterior spacetimes of mass $M$ and slightly larger mass $\widetilde{M}$, respectively. The gray region is the extension of the Penrose diagram of the heavier black hole into the past. At point $A$ an ingoing massless particle is released. At point $B$ it passes through the finite cutoff boundary in the initial black hole geometry and at $\widetilde{B}$ in the new black hole geometry. The black dashed lines are lines of constant and equal values of the dilaton.
Points $C$ and $\widetilde{C}$ are where the outgoing signal meets these lines.}
\label{fig:shockwave}
\end{center}
\end{figure}

Having derived the field theory deformation corresponding to Dirichlet boundary conditions for gravity and matter fields at finite cutoff, we would now like to understand the effect of such a cutoff on chaos in JT gravity. Specifically, we study this through the delay of an outgoing signal due to the release of matter from the boundary in the distant past, finding exponential dependence on the time separation. It may be slightly confusing that there is a time delay even though the particles do not directly interact. But this effect is well understood and illustrated in Figure \ref{fig:shockwave}. The fact that there is a time delay at all has to do with the new black hole geometry created by the infalling particle; the outgoing particle ends up closer to a horizon and therefore takes longer to escape. That this grows as one releases the infalling particle earlier in the past is simply because the outgoing particle spends even more time closer to a horizon. The particular exponential dependence simply arises from translating time near the horizon to time near the boundary; it is the usual redshift. The exponential dependence, arising from the near-horizon physics as it does, should not be affected by a finite cutoff. We   corroborate this expectation with a calculation below. 

The analysis is basically that of \cite{Engelsoy:2016xyb} adapted to a finite radial cutoff. The equations of motion of JT gravity enforce an AdS$_2$ metric
\bne 
ds^2 = - \frac{4 dX^+ dX^-}{(X^+ - X^-)^2}~,
\ene
while the metric for the black hole exterior region in Schwarzschild coordinates is given by
\bne 
ds^2 = -(r^2 - r_+^2)dt^2 + \frac{dr^2}{r^2 -r_+^2},
\ene
where $r_+$ is related to the mass of the black hole through $r_+^2 = 2\kappa^2 M$.
In Poincar\'e coordinates, the past and future horizons are located at $X^+ = 1/r_+$ and $X^- = -1/r_+$. The black hole metric is periodic in Euclidean time, corresponding to a temperature $T_{\infty} = r_+/2\pi$. The subscript here indicates that this is the temperature as measured with respect to coordinate time $t$. Later on we will have to convert this to a temperature measured by an observer at finite cutoff.

We will use two sets of Schwarzschild coordinates, with tilded quantities corresponding to the heavier black hole. We release matter from the boundary at $X^+ = X^- = 0$, which we define to be Schwarzschild time $t = \tilde t = 0$. This fixes the transformation between Poincar\'e and Schwarzschild coordinates to be
\bne
X^\pm (r,t) = \frac{1}{r_+} \frac{-\sqrt{r \mp r_+}+ e^{r_+ t}\sqrt{r\pm r_+}}{\sqrt{r \mp r_+}+ e^{r_+ t}\sqrt{r\pm r_+}}.
\ene
The dilaton is given by $\Phi(r) = r$, and the observer measuring the time delay is fixed at a constant value of the dilaton. The infalling matter of energy $\d M$ increases the mass of the black hole to $\widetilde{M} = M + \d M$. We follow in the matter until it reaches the radial cutoff surface at $r = r_c$, $t = t_1$. This is point $B$ in Figure \ref{fig:shockwave}. $t_1$ can be written in terms of $r_c$ by solving the null geodesic equation of motion, or by solving $X^- (r_c , t_1) = 0$ (and equivalently so for the point $\widetilde{B}$ in terms of the tilded quantities in the new black hole geometry). This gives 
\be\label{t1eq}
e^{r_+ t_1} = \sqrt{\frac{1+ \frac{r_+}{r_c}}{1- \frac{r_+}{ r_c}}}.
\ee

Now we consider the outgoing signal, specifically events $C$ and $\widetilde{C}$ in the figure. Event $C$ occurs at $(r_c, t_2)$ and $\widetilde{C}$ at $(\tilde{r}_c, \tilde{t}_2)$. Fixing our observer at a constant value of the dilaton sets $r_c = \tilde{r}_c$. The time $t_2$ is the time at which the observer would have measured the outgoing signal if the infalling matter had not been sent, while $\tilde{t}_2$ is the actual time measured. We want to calculate the time delay $\tilde{t}_2 - t_2$, most importantly how it depends on the time separation $t_2 - t_1$ of the ingoing and outgoing rays, as this will give us the Lyapunov exponent. Before we can do so, we first convert all the times to ones as measured by a field theory observer. Following the dictionary \cite{Hartman:2018tkw} the field theory time $t_b$ is defined through
\be
r_c^2 dt_b^2 = (r_c^2 - r_+^2)dt^2 \Rightarrow t_b = \sqrt{1-\frac{r_+^2}{r_c^2}}t,
\ee
where we took a flat field theory metric $\g_{tt} = -1$. After solving for $\tilde{t}_2$ in terms of $t_2$, by using that events $C$ and $\widetilde{C}$ occur at the same $X^+$, and converting to times and temperatures appropriate for the dual theory at finite cutoff, we obtain a time delay at large $t_{b,2} - t_{b,1}$ given by
\bne
\tilde{t}_{b,2} - t_{b,2} \approx \frac{ \d M\, \b_c}{8\pi M}\,\,\f{4\pi^2\left(1+\sqrt{1+r_c^2\b_c^2/(4\pi^2)}\right)^2}{r_c^2\b_c^2}\exp\left(\frac{2\pi}{\b_c} (t_{b,2} - t_{b,1})\right).
\ene
This is to first order in $\d M$ and nonperturbative in $r_c$. Here $\b_c = \b_{\infty}\sqrt{1-(2\pi/\b_{\infty}r_c)^2}$ is the temperature at finite cutoff. Taking a large time separation at fixed cutoff $r_c$ requires fixing $t_{b,1}$ and taking $t_{b,2}$ large, since the infalling matter is fixed to fall in from the AdS boundary at $t=0$. Notice that to this order in $\d M$ the right-hand-side can be written in terms of tilded quantities, as appropriate for the late-time observer, and still take the same functional form. Notice also that for sufficiently large $t_{b,2}$ we can ignore the $r_c$-dependent part of the prefactor by using \eqref{t1eq}. 

This bulk analysis tells us that the Lyapunov exponent of the boundary EFT is still maximal, i.e. we still have a Lyapunov exponent given by $\l_L = 2\pi/\b_c$.   In section \ref{chaosbdry} we will do a boundary quantum-mechanical calculation of the Lyapunov exponent, finding again that it remains maximal. 

\section{$T\overline{T}$ in quantum mechanics}\label{qmsec}
The particular deformation proposed in \eqref{floweq} was engineered to give physics that agrees with a finite radial cutoff for JT gravity in AdS$_2$. In this section, we study various aspects of the deformation directly in quantum mechanics, without relying on any holographically dual picture. In particular we will analyze the symmetries along the flow and show that the resulting ultraviolet action is universally given by a worldline action with nontrivial target space metric. 

\subsection{Conserved charges and supersymmetry}\label{symms}
Let us consider deformed Hamiltonians $H$ which are given as a function of the original Hamiltonian $H_0$, $H(q_i, p_i)=f(H_0(q_i,p_i))$. This captures the deformation \eqref{floweq} and others considered in this paper.  If the original Hamiltonian has a classical conserved quantity $\dot{p}_i = 0$, i.e. $\partial H_0/\partial q_i = 0$, then we have 
\be
\f{\partial H}{\partial q_i} = f'(H_0)\f{\partial H_0}{\partial q_i} = 0 \,,
\ee
 so the conserved quantity remains conserved for any analytic  function $f$. Thus all classically integrable theories remain integrable. Quantum mechanically, for a conserved charge $Q$ in the original theory we have $[H_0, Q] = 0$, which automatically gives $[f(H_0), Q] = 0$ for any analytic function $f$. Thus all quantum mechanical conserved charges remain conserved charges. 
 
It is interesting to consider the case of a theory invariant under supersymmetry. This means its Hamiltonian can be written as 
\be
H =\{ Q, Q^{\dagger}\}= Q_1^2 = Q_2^2\,,\qquad Q = \f{1}{2}(Q_1+iQ_2)
\ee
for two real supercharges $Q_i$. As we just argued, all conserved charges will remain conserved under these Hamiltonian flows, but we would like to maintain the algebra by writing $\l$-deformed supercharges which still obey the supersymmetry algebra. It is simplest to work with the real supercharges, which can be written as the square root of the Hamiltonian:
\be
H(\l) = \f{1}{4\l}\left(1-\sqrt{1-8\l Q_i^2}\right)\implies Q_i(\l) = \pm \sqrt{\f{1}{4\l}\left(1-\sqrt{1-8\l Q_i^2}\right)}\,,
\ee
with no sum over $i$ intended. The branch of the (overall) square root depends on the given eigenvalue and is chosen to match onto the eigenvalue at $\l = 0$, which can be either positive or negative. This solution arises from a flow
\be
Q_i(\l)\partial_{\l}Q_i(\l) = \f{Q_i(\l)^4}{1-4\l Q_i(\l)^2}\,.
\ee
The preservation of supersymmetry for the $T\bar{T}$ deformation in $d=2$ was considered in \cite{Baggio:2018rpv,  Chang:2018dge}.

\subsection{Worldline action}\label{uvqm}
In this subsection we solve for the deformed action under the flow 
\be
\f{\partial S_E}{\partial \l} = \int d\t  \f{T^2}{1/2-2\l T}\,.
\ee
This flow can be considered for quantum-mechanical theories with general kinetic terms. In this section we will focus on seed theories with canonical kinetic terms but otherwise arbitrary interactions. We use the approach of appendix \ref{tangent}, where we transform our deformed Hamiltonian to obtain $L = \sum_i p_i dH/dp_i - H$. Since the energy eigenvectors are unchanged under this deformation, we can write the deformed Hamiltonian as: 
\begin{align}\label{ansatznew}
H = \f{1-\sqrt{1-8\l \left(\sum_i\f{p_i^2}{2}+V(q_1,\dots,q_N)\right)}}{4\l}\,.
\end{align}
We have chosen the starting Hamiltonian to be a bosonic theory in $N$ dimensions with a general potential $V(q_1,\dots,q_N)$. The Euclidean Lagrangian is then given as 

\be\label{worldlineL}
L_E = \f{1-\sqrt{(1-4\l \dot{q}_i^2)(1-8\l V(q_1,\dots, q_N))}}{4\l}\,.
\ee
This form of the Lagrangian can be interpreted as a worldline action with a nontrivial target space metric. To see this, write \eqref{worldlineL} as
\be\label{worldlineS}
S_E =  \f{1}{4\l} \int d\tau \left(1-\sqrt{g_{\m\n}\dot{X}^\m\dot{X}^\n }\right),
\ee
where $g_{\m\n} = \d_{\m\n}(1-8\l V(X^1,\dots,X^N))$ is a curved target space metric. Picking static gauge $X^0(\t) = \t$, $X^i(\t)=2\sqrt{-\l} q_i(\t)$ gives the action from before. The potential $V(X^1,\dots,X^N)\defeq V(q_1(X^1),\dots, q_N(X^N))$, and the metric becomes trivial for vanishing potential. The mass is identified as $m=1$ since the expansion at small velocities gives a kinetic term $\f{1}{2}\dot{q}^2$.
Thus the one-dimensional particle is embedded in an $(N+1)$-dimensional Euclidean space. This is analogous to the $T\overline{T}$ deformation of free bosons in $d=2$ leading to a Nambu-Goto action \cite{Cavaglia:2016oda}. In that case, however, adding interactions seems to ruin the Nambu-Goto form of the UV action \cite{Bonelli:2018kik}, while in our case interactions simply change the target space metric while maintaining the worldline interpretation. (For a deformation closely related to $T\overline{T}$ that modifies interacting bosons in $d=2$ to Nambu-Goto with a nontrivial target space metric, see Appendix \ref{ccharges}.)  For $\l > 0$ the worldline interpretation requires either wrong-sign kinetic terms or a Lorentzian target space metric, while for $\l < 0$ the interpretation is standard. 
This analysis can be generalized to a seed theory with nontrivial metric $h_{ij}\dot{q}^i \dot{q}^j$. 

We obtained the Lagrangian \eqref{worldlineL} from \eqref{worldlineS} by fixing a particular gauge. Starting instead from \eqref{worldlineL}, there are in fact different ways to gauge unfix it and obtain different interpretations of the deformed theory. For example, one other very fruitful way of gauge unfixing \eqref{worldlineL} is by interpreting the shift $1/4\l$ as coming from a Chern-Simons term for a worldline gauge field $q\int d\tau a$. 
We can complete this worldline gauge field into a target space gauge field $A_\mu$. Parameterizing our worldline by target space coordinates $X^\mu(\t)$, we have
\be
S_E = \int d\t \left(\frac{1}{2e} \dot{X}^{\mu}\dot{X}_{\mu} + \frac{1}{2}e m^2 \right) - i  q \int d\tau A_{\mu} \dot{X}^{\mu}
\ee
with $e$ an einbein, $ds^2 = e^2 d\t^2$. Solving the metric equation of motion for $e$ and plugging it back into the action, we find
\be
S_E = m\int d\tau \sqrt{\dot{X}^{\mu}\dot{X}_{\mu}} - i q \int d\tau A_{\mu}\dot{X}^{\mu}.
\ee
For $\l < 0$, choosing static gauge ($\dot{X}^0 = 1$) and setting $A_{\mu}dX^{\mu} = (i/4\l) d\t$ and $m = 1/4\l$, we find the same action as before. On the other hand, for $\l >0$ we need to use the other branch of the solution to the equation of motion of $e$. The other identifications remain the same (given the target space metric $g_{\mu\nu} = -\eta_{\mu\nu}(1-8\l V)$). 
  Interpreting the deformed theory as a worldline coupled to a gauge field bears some resemblance with recent proposals of Kitaev and Suh \cite{Kitaev:2018wpr} and Yang \cite{Yang:2018gdb}, where quantum JT gravity is interpreted as a worldline in AdS$_2$ with an electric field.

Fermionic parameters can also be treated. As discussed in appendix \ref{tangent}, the flows are easier in this case since the kinetic term does not flow. We consider an arbitrary seed theory of $N$ fermions with canonical kinetic terms
\be
S_E = \int d\tau \, \left(\frac{1}{2}\; \overline{\psi}_{i}\dot{\psi}^{i} + V\left(\overline{\psi}_{i},\psi_{i}\right)\right).
\ee
As usual the deformed action can be found by Legendre transorming the deformed Hamiltonian. The result is
\be
S_E =\int d\tau \left( \frac{1}{2}\,\overline{\psi}_{i}\dot{\psi}^{i} + \frac{1}{4\l}\left(1 - \sqrt{1-8\l V(\overline{\psi}_{i},\psi_{i})} \right)\right).
\ee
To see that this can be written as a worldline action we need to re-introduce the bosonic parameters. For a seed theory 
\be
S_E = \int d\tau \, \left(\frac{1}{2}\dot{q_i}^2+\frac{1}{2}\; \overline{\psi}_{i}\dot{\psi}^{i} + V\left(q_i,\overline{\psi}_{i},\psi_{i}\right)\right)
\ee
we find a deformed action
\begin{align}
S_E &=\int d\tau \left( \frac{1}{2}\,\overline{\psi}_{i}\dot{\psi}^{i} + \frac{1}{4\l}\left(1 - \sqrt{(1-4\l \dot{q}_i^2)(1-8\l V(q_i, \overline{\psi}_i,\psi_i))} \right)\right)
\\&=\int d\tau \left( \frac{1}{2}\,\overline{\psi}_{i}\dot{\psi}^{i} +\f{1}{4\l}\left(1-\sqrt{g_{\m\n}\dot{X}^\m \dot{X}^\n}\right)\right),
\end{align}
where the various parameters are defined as before. 

To see that this is the appropriate notion of a worldline action, let us recall what happens when we couple the seed theory directly to an einbein:
\be
S_E = \int d\tau \, \left(\frac{1}{2e}\dot{q}_i^2+\frac{1}{2}\; \overline{\psi}_{i}\dot{\psi}^{i} + eV\left(q_i,\overline{\psi}_{i},\psi_{i}\right)\right).
\ee
The equation of motion for the einbein $e$ gives $e = \sqrt{\f{\dot{q}_i^2}{2V}}$, which upon inserting back into the action gives 
\be
S_E = \int d\t \, \left(\f{1}{2} \; \overline{\psi}_{i}\dot{\psi}^{i}+\sqrt{2V \dot{q}_i^2}\right)=\int d\t \, \left(\f{1}{2} \; \overline{\psi}_{i}\dot{\psi}^{i}+\sqrt{g_{ij}\dot{q}^i\dot{q}^j}\right),\qquad g_{ij} = 2\d_{ij}V\,.
\ee
So we see that the fermion kinetic term is invariant under einbein coupling, just as it is invariant under our $T\overline{T}$ deformation. This is because it is topological and does not contribute to the stress tensor, which is necessary to couple to an einbein. 

It is also interesting to study more general initial theories containing higher derivative terms. The phase space of such theories is larger and requires more canonical momenta and coordinates than the usual pair $\{p,q\}$ per field we have been considering thus far. These momenta are obtained via the Ostrogradsky formalism \cite{Woodard:2015zca}. The flow of such higher derivative theories is then obtained in the same way as ordinary theories considered in this section. An interesting example of such a higher-derivative theory is the Schwarzian theory, which we will consider in section \ref{schwarziansec}. 

\subsection{Thermodynamics}\label{thermo}
It is interesting to consider the finite-temperature partition function of the deformed theory, which can be written as 
\be
Z(\b)_\l = \int_{-\infty}^\infty dE\, e^{-\b f(E)}\rho(E)\,.
\ee
For now we will consider general invertible $f(E)$, restricting to the one relevant to our $T\overline{T}$ deformation at the end. Performing a change of variables $E \rightarrow f^{-1}(E)$ lets us write this as
\be
Z(\b)_\l = \int dE \,e^{-\b E}\rho\left(f^{-1}(E)\right) \f{d f^{-1}(E)}{dE}\,.
\ee
This means the deformed finite-temperature partition function $Z(\b)_\l = \int dE \,e^{-\b E}\rho_\l (E)$ has  density of states 
\be\label{deformedrho}
\rho_\l (E) = \rho\left(f^{-1}(E)\right) \f{d f^{-1}(E)}{dE}\,.
\ee
For certain $f(E)$, $Z(\b)_\l$ can be written as an integral transform of the partition function of the seed theory $Z(\b)$:
\be
e^{-\b f(E)} = \int_0^\infty d\b' e^{-\b' E}K(\b, \b')\implies Z(\b)_\l = \int_0^\infty d\b' Z(\b') K(\b, \b')\,.
\ee
The kernel is simply the inverse Laplace transform of the Boltzmann weight of the deformed theory:
\be
K(\b, \b') =\f{1}{2\pi i} \int_{ - i \infty}^{+i\infty} dE\, e^{-\b f(E)+\b' E}\,.
\ee
Thus for certain $f(E)$ for which this integral converges, we can write the deformed partition function as an integral transform of the original partition function. For the $T\overline{T}$ deformation $f(E) = \f{1}{4\l}(1-\sqrt{1-8E \l})$ with $\l < 0$, we have
\be\label{integraltransform}
Z(\b)_\l = \frac{\b}{\sqrt{-8\pi \l}}\int_0^\infty \frac{d\b'}{\b'^{3/2}}\, \exp\left(\f{(\b-\b')^2}{8\b' \l}\right) Z(\b')\,.
\ee
Depending on the spectrum of the quantum mechanical theory, this relation between partition functions may break down at some value of $\beta$. In particular, consider a theory with a linear specific heat, $\log Z = c T$. The exponential suppression from the kernel at small $\b'$ then competes with the exponential divergence of $Z(\b')$. The critical temperature at which $Z(\b)_\l$ diverges is $T = (-8\l c)^{-1/2}$. This is a standard Hagedorn divergence, which can be seen by looking at the entropy at high energy. The seed theory has $S(E_i) = (1-\b \partial_\b)\log Z(\b) = 2 \sqrt{c E_i}$. Due to the quadratic relation  between the energy of the seed theory and the deformed theory, $E_i = E_\l - 2 \l E_\l^2$, this leads to a Hagedorn density of states $\exp\left( E_\l \sqrt{-8 \l c}\right)$, from which the Hagedorn temperature $T = (-8\l c)^{-1/2}$ can be read off. Picking a seed theory with a faster density of states at high energy can lower the Hagedorn temperature and push it all the way to zero. Of course, these formulas are also valid for quantum-mechanical theories with a finite spectrum, for which no divergences can arise. 


\section{Flowing the Schwarzian theory}\label{schwarziansec}
In this section we would like to apply some of the techniques of the previous sections to the Schwarzian theory. This is a theory which captures, among other things, Jackiw-Teitelboim gravity in AdS$_2$\cite{Maldacena:2016upp, Jensen:2016pah, Engelsoy:2016xyb}.
\subsection{Deformed Schwarzian action}
\label{sec:DSA}
Picking a metric $(r^2-1)d\t^2+\f{dr^2}{r^2-1} $, one introduces an infrared cutoff on the spacetime with some curve $(\t(u),r(u))$, with $u\sim u+2\pi$ the boundary time coordinate. This cutoff problem is formulated slightly differently than in the $T\overline{T}$ literature, where a sharp radial cutoff $r=r_c$ is chosen. Instead, one fixes the length of the cutoff curve with a Dirichlet condition on the metric $ds^2\big|_{\text{bdry}} = \L^2 du^2$. At leading order in $\L$, preserving this boundary condition on the metric means picking $r(u) = \L/\t'(u)$. In particular, the length of the curve remains fixed under diffeomorphisms that would change a sharp radial cutoff. This means that we need to slightly reformulate the problem of the flow of the bulk on-shell action. Instead of flowing $\log Z[g_{ij}^0(r_c, \t), \phi^0(r_c, \t)]$ with respect to $r_c$, we need to flow $\log Z[g_{ij}^0(\L, u), \phi^0(\L,u)]$ with respect to $\L$. This can be done analogously to section \ref{general2d}, and the final result is the same up to $r_c \rightarrow \L$. Thus the boundary flow is identical and the map with the bulk is modified to be $\l = 2\pi G/(\Phi_r \L^2)$. 

With a cutoff curve $(\t(u),r(u))$, the bulk theory reduces to an equivalent boundary theory at leading order in $\L$, with action given by 
\be\label{schwarzian}
S_E = -C \int du \,  \left(\text{Sch}(\t,u)+\t'^2/2\right)\,,\qquad C = \f{\Phi_r}{8\pi G}\,,
\ee 
where $\Phi_r$ is the renormalized value of the dilaton, which we have taken to be a constant, and Sch$(\t,u)$ is the Schwarzian
\be
\text{Sch}(\t,u) = \left(\frac{\t''}{\t'}\right)'- \frac{1}{2}\left(\frac{\t''}{\t'}\right)^2.
\ee
 This is found by evaluating the boundary action (corresponding to the Gibbons-Hawking-York extrinsic curvature term and a counterterm) for a particular curve. The stress tensor of this theory is given as $T = C (\text{Sch}(\t,u)+ \t'^2/2)$. The action has corrections in $\L$ which we will consider in section \ref{bulkcomp}. 
 
We now want to flow the Schwarzian action under \eqref{floweq}. To apply the technique of subsection \ref{uvqm}, we first compute the Hamiltonian of the Schwarzian theory in canonical variables. Dropping the total derivative term, the canonical coordinates are $q_1 = \t$ and $q_2 = \t'$. The momenta are determined using the Ostrogradsky formalism \cite{Woodard:2015zca},
\begin{align}
p_1 &= \frac{\partial L}{\partial \t'} - \frac{d}{du}\left(\frac{\partial L}{\partial \t''}\right) = C \left( \frac{\t''^2}{\t'^3} - \frac{\t'''}{\t'^2} - \t' \right),\\
p_2 &= \frac{\partial L}{\partial \t''} = C \frac{\t''}{\t'^2}\,.
\end{align}
The undeformed and deformed Hamiltonian then become
\be
H_0 = \frac{p_2^2 q_2^2}{2C} + \frac{C}{2}q_2^2 + p_1 q_2\,,\qquad H(\l) = \frac{1}{4\l}\left( 1 - \sqrt{1-8\l H_0} \right).
\ee
Legendre transforming this to a Lagrangian and analytically continuing to Euclidean signature through $u \to -i u, q_2 \to i q_2$, gives 
\be\label{LSchdef}
L_E(\l) = - \frac{(\t' - e^{\phi})^2}{8\l \t' e^{\phi}} + \frac{C}{2} \frac{e^{\phi}}{\t'} (\phi'^2 - \t'^2)\,,
\ee
where we have substituted $q_1 = \t$ and $q_2 = e^{\phi}$. These substitutions are so the resulting Lagrangian is identical to the one obtained by deforming the Schwarzian theory in Liouville variables $\t' = e^{\phi}$. The starting theory in that case is given by $L_E = C(\phi'^2 - e^{2\phi})/2+\omega (\t' - e^{\phi})$, where $\omega$ serves as a Lagrange multiplier enforcing the constraint $\t' = e^{\phi}$.

A few comments about \eqref{LSchdef} are in order. First, the $\l \to 0$ limit seems to be ill-defined here, but we need to remember that at $\l = 0$ the momentum $p_1$ appears linearly, enforcing the constraint $\t' = e^{\phi}$ in the variables chosen above. Plugging this into \eqref{LSchdef} as we take $\l\rightarrow 0$ gives us the Schwarzian theory, and there are then corrections to this in $\l$. We will encounter a similar phenomenon in section \ref{nonpert} and show precisely that there are no issues at the level of the full path integral. Second, it can be checked that the Lagrangian \eqref{LSchdef}, together with its stress tensor 
\be
T(\l) = \frac{1}{4\l}\left(1 - \frac{e^{\phi}}{\t'}\right),
\ee
satisfy the flow equation \eqref{floweq}. Finally, upon substituting the solution to the $\t$ equation of motion that at $\l \to 0$ gives $e^{\phi} = \t'$, we find
\be
L_E(\l) = \frac{1}{4\l}\left( 1 - \sqrt{\left(1 - 4 \l C \phi'^2\right)\left( 1+ 4 \l C e^{2\phi}\right)} \right).
\ee
This is precisely the action one would get from flowing the Liouville quantum mechanics $L_E = C(\phi'^2-e^{2\phi})/2$. This Lagrangian can be obtained by the substitution $\t' = e^{\phi}$ in the Schwarzian Lagrangian, although as seen below \eqref{LSchdef} for a precise equivalence between the two theories we need to incorporate a Lagrange multiplier enforcing the constraint $\t' = e^{\phi}$.

\subsection{Comparison to bulk}\label{bulkcomp}
Another way to move the cutoff surface into the bulk is to perform the procedure of \cite{Maldacena:2016upp} to higher orders. We can compute corrections to the boundary term in the bulk action systematically. Working in the finite temperature geometry $ds^2 = (r^2-1)d\t^2+dr^2/(r^2-1)$, we pick a wiggly cutoff $(r(u),\t(u))$ with boundary conditions $ds^2|_{\text{bdry}} = \Lambda^2 du^2$ and $\Phi|_{\text{bdry}} = \Lambda \Phi_r$ for $\L$ large. The extrinsic curvature of this cutoff surface is given as 
\be\label{extr}
K = \sqrt{r^2-1}\left(\frac{\t' \left(r''  + r (3 {r'}^2 + (r^2 -1)^2 {\t'}^2 - r r'')\right) + (r^2 - 1)r' \t''}{\left({r'}^2+\left(r^2-1\right)^2 {\t'}^2\right)^{3/2}}\right) 
\,.
\ee
Using the boundary condition to expand $r(u)$ in terms of $\t(u)$ gives 
\be\label{rexp}
r = \frac{\Lambda }{\tau'}+\frac{{\t'}^4-{\t''}^2}{2 \Lambda  {\t'}^3} + \mathcal{O}(\L^{-3})\,.
\ee
The bulk action vanishes due to the path integral over the dilaton. The boundary action $-(8\pi G)^{-1}\int du \Lambda^2 \Phi_r (K-1)$ can be obtained by plugging \eqref{rexp} into $\eqref{extr}$ and expanding in $\L$. This gives
\begin{align}\label{schwaction2}
S_E = -\f{1}{8\pi G} \int du\, \Phi_r &\left(\text{Sch}(\t,u) + \frac{{\t'}^2}{2}-\frac{{\tau'}^7+\tau''' {\tau''}^2+6 \tau'''{\tau'}^4}{8 {\tau'}^3\L^2}\right.\nonumber\\
&\left.\quad\quad +\left(\frac{\tau'' \left(-9 {\tau''}^2+2 {\tau'}^4+8 \tau'''\tau'\right)}{8 {\tau'}^3\L^2}\right)' + \mathcal{O}(\Lambda^{-4})\right).
\end{align}
So we can now check if the corrections to the Schwarzian action computed in this way agree with the ones computed by flowing the action via \eqref{floweq}.
The full nonperturbative action with respect to this flow, for constant dilaton $\Phi_r = 8\pi G C$, is given in \eqref{LSchdef}. Since we will only illustrate the comparison at first order in $\l$ we can simply consider the single-field Lagrangian $L(\l) = L_0+2\l L_0^2 + \mathcal{O}(\l^2)$, where we used the fact that the stress tensor of the finite-temperature Schwarzian action is (minus) the Schwarzian. Notice also that for constant dilaton the total derivative term can be dropped, which leads to a much simpler action.

Before comparing, we need to recall that our flow is written for the on-shell action in the bulk. We have applied the flow to the Schwarzian action, which is an off-shell action in the sense that the bulk metric equations of motion have not been imposed. We should therefore not expect that the resulting action should agree with the corrections to the Schwarzian obtained by the procedure above. Instead, we should expect agreement on-shell, i.e. upon imposing the metric equations of motion. These equations are 
\be
\left(g_{\m\n}\nabla^2-\nabla_\m\nabla_\n\right)\Phi - g_{\m\n}\Phi = 0
\ee
and are solved by 
\be\label{phisol}
\Phi = \a \,r + \sqrt{r^2-1} \left(\b \sin \t+\g \cos \t\right).
\ee
The equation of motion of the Schwarzian theory is obtained by varying the Schwarzian action with respect to $\t(u)$. This can be interpreted as an equation for $\Phi_r =  \Phi/\L$, and it is solved by \eqref{phisol} above, after rewriting it purely in terms of $\t$ through the expansion \eqref{rexp} at leading order. It can be checked that this equivalence holds to higher orders in $\L$; for the case at hand it means that the equation of motion following from \eqref{schwaction2} is solved, to order $\L^2$, by
\be
\Phi_r =\frac{\alpha +\beta  \sin \tau +\gamma  \cos \tau }{\tau '}+ \frac{\alpha  \tau '^4-\tau ''^2 (\alpha +\beta  \sin \tau +\gamma  \cos \tau )}{2 \Lambda ^2 \tau '^3}+\mathcal{O}(\L^{-4})\,.
\ee
Now we see that imposing the bulk metric equations of motion to go on-shell can instead be done by imposing the equation of motion of \eqref{schwaction2}. The saddle $\t(u) = u$ remains a saddle of the corrected Schwarzian action, so we can compare the actions on this saddle. We immediately see that the only term that contributes in the $\mathcal{O}(\L^{-2})$ part of the action is the $\t'^4/(8\L^2)$ piece. Using $\l = 2\pi G/(\Phi_r \L^2)$ shows that the action obtained by our flow agrees with the action \eqref{schwaction2} when both are evaluated on shell. 

\subsection{Chaos in the deformed Schwarzian theory}\label{chaosbdry}
In this section we investigate how a finite cutoff affects the maximally chaotic behaviour of JT gravity, by directly calculating the Lyapunov exponent of the deformed boundary action. 
\subsubsection{Propagator of quantum fluctuations}\label{sec:fluc}
 We begin by looking at the fluctuations around a saddle and computing their two-point function. This will be an important piece of the calculation in section \ref{sec:OTOC}, where we will compute the gravitational corrections to an out-of-time-order four-point function. 

We want to find a saddle of the deformed Schwarzian theory \eqref{LSchdef}. The undeformed saddle $\tau(u) = u$ should remain a saddle, since the deformed theory (thought in terms of the single variable $\tau$ and not both $\tau$ and $\phi$) has only derivatives of $\tau$ appearing in the action. Setting $\tau(u) = u$ and solving the remaining equation of motion gives
\be
e^{-\phi(u)} = \sqrt{1+ 4 \l C }\,.
\ee
We expand the action \eqref{LSchdef} to quadratic order in the fluctuations $\t = u+\e(u)$, $e^{\phi} = e^{\eta(u)}/\sqrt{1+4\l C}$ with $\e(u+2\pi) = \e(u)$ and $\eta(u+2\pi) = \eta(u)$. Ignoring a constant piece, we have
\be
S_E(\l) = \f{1}{8\l \sqrt{1+4\l C}}\int_0^{2\pi} du \left(-\e'(u)^2 + 4 \l C \eta'(u)^2 - (1+ 4 \l C)\eta(u)^2 + 2 \eta(u) \e'(u) \right).
\ee
Expanding in Fourier modes
\be
\e(u)=\sum_{n\in \mathbb{Z}} \e_n e^{i nu}, \quad \eta(u) = \sum_{n\in \mathbb{Z}} \eta_n e^{i nu},
\ee
the action becomes
\be\label{quadaction}
S_E(\l) = \frac{1}{2} \sum_{n \in \mathbb{Z}}  \xi_n^i A_{ij} \xi_{-n}^j\,, \quad A = \frac{\pi}{2\l\sqrt{1+4 \l C}} \begin{pmatrix}
-n^2 & in \\
-in & -1 + 4 \l C (n^2 - 1)
\end{pmatrix}\,,
\ee
where $\xi_n = (\e_n \;\, \eta_n  )$. Just like the undeformed Schwarzian action, the above kernel is not invertible for $n=0, \pm 1$, originating from the unbroken $SL(2,\mathbb{R})$ gauge symmetry. The relevant correlator that we want to extract is the one for the fluctuation around the $\t$ saddle, since $\t$ couples to bulk matter as we discuss below. Computing the generating functional of $\e$ and $\eta$ correlators, which amounts to inverting $A$, we find that the $\e$ two-point function is 
\be
\langle \e(u) \e(0)\rangle = \frac{\sqrt{1+4\l C}}{2\pi C} \sum_{n\neq 0,\pm 1} \frac{1 - 4 \l C (n^2-1)}{n^2(n^2 -1)}e^{i n u}.
\ee
Note that this correlator goes to the undeformed one as $\l \to 0$. 
This sum can be done explicitly in terms of special functions, but it is more transparent to perform it for $0 < u < 2\pi$ by writing it as a contour integral:
\be
\langle \e(u) \e(0)\rangle = \frac{\sqrt{1+4\l C}}{2\pi C} \oint_\mathcal{C} \f{ds}{e^{2\pi i s}-1}\f{(1+4\l C(1-s^2))}{s^2(s^2-1)}e^{i su}\,.
\ee
The contour initially circles all integer $ s \neq 0, \pm 1$, but this is deformed to contours circling $s  = 0, \pm 1$ and one at infinity, which can be dropped. The correlator evaluates to 
\begin{align}\label{gravcorr}
\langle \e(u) \e(0)\rangle &= \frac{\sqrt{1+4 \l C}}{2\pi C}\left( 1 - (1+4\l C)\left(\frac{\pi^2}{3} - \pi u + \frac{u^2}{2}\right)\right. \nonumber\\
&\quad\quad\left. + \left(\frac{5}{2} + 8 \l C\right)\cos u + (u-\pi)\sin u \right).
\end{align}
This does not exhibit $2\pi$-periodicity because of the assumption that allowed us to drop the contour at infinity; the full answer obtained by doing the sum explicitly is simply the above repeated over $2\pi$ intervals, which can be represented in terms of special functions.

\subsubsection{Out-of-time-order four-point function}
\label{sec:OTOC}
We want to compute an out-of-time-order four-point function of an operator dual to a bulk matter field. The theory at finite cutoff is determined by flowing the Schwarzian coupled to the matter sector. Like in the previous section we will consider a massless scalar in the bulk. The flow of such a theory was considered in \ref{general2d}, and we have
\begin{equation}
\tilde{t}^r_r = \frac{1}{2}\left(\frac{\pi_\chi^2}{g^0} - g^{0ij}\del_i \chi \del_j \chi \right).
\end{equation}
For a bulk metric of the form $ds^2 = N(r)^2dr^2 + r^2 \g_{\t\t} d\t^2$ and identifications $\pi_\chi (r_c, \t)=\sqrt{\g}\mathcal{O}_\chi (\t)$, $\chi(r_c, \t) = J_\chi(\t)$, $r_c^{-1} = \sqrt{4\l C}$, and $g^0_{\t\t} = r_c^2 \g_{\t\t}$, the flow equation becomes 
\begin{equation} \label{eqn:flow}
\frac{\del S_E}{\del \l} = \int d\tau \sqrt{\gamma}\, \frac{T^2 - \frac{1}{4} \sqrt{C/\lambda}\left(\mathcal{O}_\chi^2- (\partial J_\chi)^2\right) }{\frac{1}{2}- 2\lambda T} \,.
\end{equation}
While this is the exact flow that needs to be considered to match onto a finite Dirichlet cutoff in the bulk, we will only consider the corrections coming from the gravitational sector. We expect these to be the important pieces for the part of the OTOC that grows at late times. The deformed boundary action we consider is given by
\bne
S_E(\lambda) = S_{\text{matter on-shell}} +\int du \,L_E(\l)\,,
\ene
with $L_E(\l)$ given by \eqref{LSchdef}.
The on-shell matter action is found by solving the bulk equations of motion for given boundary conditions $\chi_r$. It couples to gravity through the boundary degree of freedom $\t(u)$ in the following way,
\bne \label{eqn:matter} S_{\text{matter on-shell}} = - D \int du du' \left [ \frac{\t'(u) \t'(u')}{(\t(u)- \t(u'))^2}\right]^\Delta \chi_r (u) \chi_r (u'),\quad D = \frac{(\Delta - \frac{1}{2})\Gamma(\Delta)}{\sqrt{\pi}\Gamma(\Delta - \frac{1}{2})}\,, \ene
where on the boundary $\chi_r (u)$ acts as a source for an operator of scaling dimension $\Delta = \frac{1}{2}(1 + \sqrt{1+ 4m^2})$. We take $\chi$ to be massless, giving $\Delta =1$ and $D = 1/2\pi$.

We turn now to how the Lyapunov exponent is extracted from the analytic continuation of the Euclidean 4-pt function
\bne F = \frac{\< V(u_1) V(u_2) W(u_3) W(u_4) \rangle - \< V(u_1) V(u_2) \rangle \< W(u_3) W(u_4) \rangle}{\< V(u_1) V(u_2) \rangle \< W(u_3) W(u_4) \rangle}.
\ene
In our case, this can be written as a 2-pt function of an operator $B(u_1, u_2)$:
\bne F = \< B(u_1, u_2 ) B(u_3, u_4)\rangle. \ene
$B(u_1, u_2)$ is found by expanding the on-shell matter action \eqref{eqn:matter} about the classical saddle (setting $\beta = 2\pi$),
\bne \t(u) = \tan \left ( \frac{u+ \e (u)}{2} \right ), \ene
to linear order in the fluctuation parameter $\e (u)$, giving
\bne \label{eqn:B} B(u_1, u_2) = \Delta \left(\e'(u_1) + \e' (u_2) - \frac{\e (u_1) - \e (u_2)}{\tan \frac{u_{12}}{2}}\right). 
\ene
It is clear that since the four-point function is a two-point function of $B$, which in turn is linear in $\e$ and its derivatives, we will need to calculate the propagator $\< \e (u) \e (0) \rangle $. This was done in section \ref{sec:fluc} and the final answer is given in \eqref{gravcorr}. Plugging this into the equation for $B$ given by \eqref{eqn:B} gives us the four-point function. To extract the Lyapunov exponent we Wick rotate to Lorentzian time $u \to i \hat u$ and look at the late time behaviour. After restoring $\beta$, in this limit we find the OTOC to be
\bne 
F_{VWVW} \sim \frac{\beta}{C} \sqrt{1+ \l C\frac{16\pi^2}{\b^2} }\,\exp\left(\frac{2\pi}{\beta} \hat{u}\right) .
\ene
Thus the Lyapunov exponent is unaffected, in agreement with the bulk computation of subsection \ref{shocksbulk}.\footnote{Precisely this same answer is obtained if one considers the theory $L_E(\l) = \f{1}{4\l}(1-\sqrt{1+8\l C(\text{Sch}(\t,u)+\t'^2/2)})$ but ignores nonperturbative contributions.} 
%

\subsection{Thermodynamics}\label{thermosch}
Let us consider the finite temperature partition function of the deformed Schwarzian theory using the techniques developed in subsection \ref{thermo}. The undeformed partition function is one-loop exact \cite{Stanford:2017thb} and given by 
\be
Z(\b) = \frac{\a}{\b^{3/2}} \exp\left(\frac{2\pi^2 C}{\b}\right),
\ee
where $\a$ is a dimensionful constant. The associated density of states is given by
\be
\rho(E) = \frac{\a}{\sqrt{2\pi^3 C}} \sinh\left(2\pi \sqrt{2C E}\right).
\ee
The deformed theory thus has a density of states given by \eqref{deformedrho}:
\be\label{schwdens}
\rho_\l(E) = \frac{\a}{\sqrt{2\pi^3 C}} (1-4\l E) \sinh\left(2\pi \sqrt{2C E(1-2\l E)}\right).
\ee
It would be interesting to investigate a matrix model description of such a density of states, especially for $\l > 0$ which -- if one were to truncate the spectrum where it becomes complex -- pulls the theory away from the double-scaled limit studied in \cite{Saad:2019lba}.

Using the integral transform \eqref{integraltransform}, we can find the exact deformed partition function for $\l < 0$:
\be
Z_{\l}(\b) = \frac{\a\,\b\,e^{-\frac{\b}{4\l}}}{\sqrt{-2\pi \l}(\b^2 + 16\pi^2 C \l)} K_2\left(-\frac{1}{4\l}\sqrt{\b^2 + 16\pi^2 C \l}\right),
\ee
where $K_2(z)$ is the modified Bessel function of the second kind. This partition function exhibits a Hagedorn divergence at $\b = 4\pi\sqrt{-C \l}$, which is precisely the value at which the linearized action in the previous subsection diverged, but is otherwise well-defined. In particular it can be analytically continued into the regime of $\l > 0$, which is appropriate for JT gravity at finite cutoff. The partition function will then be complex.

\section{$T\bar T$ as quantum mechanics coupled to gravity}\label{nonpert}

In this section we propose a nonperturbative definition of quantum-mechanical theories deformed by our flow \eqref{floweq}. It proceeds analogously to the description of the $T\overline{T}$ flow as a coupling to JT gravity \cite{Dubovsky:2017cnj, Dubovsky:2018bmo}. In our case, we propose that the deformed quantum mechanics is equivalent to coupling the initial (undeformed) theory to a theory of one-dimensional gravity. We will perform the path integral over the additional fields introduced in the one-dimensional gravitational theory, which will be an einbein and a compact scalar, after which we will see that we recover \eqref{integraltransform}. We work in Euclidean signature throughout.

The coupling we propose is
\be\label{coupdef}
Z_{\l}(\b) = \int \frac{\mathcal{D}e\mathcal{D}X\mathcal{D}\Phi}{\rm Vol(\rm Diff)} e^{-S_0[e,\Phi]-S[e,X;\l]  }\,,
\ee
with $S_0[e,\Phi]$ the undeformed theory constructed out of fields $\Phi(\t)$ and placed on a one-dimensional metric with einbein $e(\t)$. The coordinate $\t$ is compact: $\t \sim \t + \b'$. We have divided by the volume of the group of time reparametrizations. The reparameterization-invariant action $S[e,X;\l]$ is
\be\label{actionS}
S[e,X;\l] = -\frac{1}{8\l} \int_0^{\b'}e\,d\t \left(e^{-1}\dot{X} -1\right)^2 = -\frac{1}{8\l} \int_0^{\b'}d\t\, \left(e^{-1}\dot{X}^2 - 2 \dot{X} + e\right),
\ee
where $\dot{X} \equiv \partial_\t X$. The field $X$ is compact with radius $\b$ and thus satisfies $X(\t + \b') = X(\t) + m \b$ with $m$ the winding number around the target space circle. 

We want to fix a gauge where $e = 1$, keeping $\tau\sim\tau+\b'$. This fixes all reparameterizations except constant shifts $\t\rightarrow \t + c$ and time reversal $\t\rightarrow -\t$. These residual symmetries are analogous to the conformal Killing group in the string theory worldsheet path integral. We will divide out by the volume of the group of these residual symmetries explicitly in the end. The einbein gauge fixing can be done using the Faddeev-Popov procedure. The path integral over einbeins will then reduce to an ordinary integral over $\b'$. We define the Faddeev-Popov measure as
\be\label{defmeas}
1=\D_{FP}(e)\int_0^\infty d\b' \int \mathcal{D}\z \,\d(e-1\,^\z)\,,
\ee
where $\z$ is our diffeomorphism and $1\,^\z$ a diffeomorphism of the fiducial einbein $e=1$. A transformation that is simultaneously a small diffeomorphism and a small change in this einbein (the latter interpreted as a change in the modular parameter $\b'$) gives $\delta e = \dot{\z}+ \d \b'/\b'$.\footnote{The second term in $\d e$ is obtained as follows. The metric is $dt^2$ and by defining $t = \b' x$, we can make the $\b'$ dependence explicit. Deforming $\b'$ to $\b' + \d \b'$ yields a deformed metric, which to first order in $\d \b'$  is $(\b'^2 + 2 \b'\d \b') dx^2$. Now, going back to the original $t$ coordinate, the change in the einbein due to a change in the modular parameter of the circle is $\d e = \d \b'/\b'$.} Writing the delta function as a Fourier integral by introducing a field $\w$, expanding it for a small transformation around unit einbein, and trading $\d \b'$, $\w$ and $\z$ for Grassmann fields $a$, $b$, and $c$ to invert the measure gives 
\be
\D_{FP}(e=1) = \int_0^\infty d\b' \int \mathcal{D}c \mathcal{D}b \mathcal{D} a \,\exp\left(-4\int_0^{\b'} d\t \left(b\dot{c} -  a b/\b' \right)\right).
\ee
The constant prefactor on the ghost action comes from a particular normalization of the fields. Performing the path integral over $a$ and inserting \eqref{defmeas} into \eqref{coupdef} gives
\be
Z_\l(\b) = \int_0^\infty d\b'\int \mathcal{D}c \mathcal{D} b\mathcal{D}\Phi \mathcal{D} X \left(\f{4}{\b'}\int_0^{\b'} d\t \,b\right)\, e^{-S_0[e=1, \Phi]-S[e=1, X; \l]-4\int_0^{\b'} d\t b \dot{c}}\,,
\ee
We ended with one $b$ ghost insertion, matching the number of moduli of our manifold, which is the expected answer. Before we evaluate the ghost piece, let us go back and first compute the path integral over $X$. We expand the field $X$ as 
\be\label{modes}
X(\t) = \t \left(\frac{m \b}{\b'}\right) + \frac{1}{\sqrt{\b'}} \sum_{n\in \mathbb{Z}} e^{2\pi i n \t / \b'} q_n~,
\ee
with $q_n = q_{-n}^*$. The first term is a solution to the equation of motion subject to the winding $m$ boundary condition, and the sum represents fluctuations around that saddle. Due to the compactness of $X$ we have $q_0 \sim q_0+\beta \sqrt{\b'}$. The action splits into a winding part and a fluctuation part. The winding part can be evaluated straightforwardly:
\be
S_m[e,X;\l] = -\frac{1}{8\b'\l} \left(m\b - \b'\right)^2.
\ee
The fluctuating part is also not too complicated. Let us first do the non-zero-mode piece. This gives 
\be
\int \mathcal{D}q\,e^{-S[e=1,q]} = \left(\det\left(\frac{1}{8\pi \l}\partial_{\t}^2\right)\right)^{-1/2} = \prod_{n>0} \left(\frac{-2\l \b'^2}{\pi n^2}\right),
\ee
where we used the eigenvalues $-(2\pi n)^2/\b'^2$ of the differential operator $\partial_{\t}^2$. This product can be evaluated using zeta function regularization and yields
\be
\prod_{n>0} \left(\frac{-2\l \b'^2}{\pi n^2}\right) = \frac{1}{\b'\sqrt{-8\pi \l}}\,.
\ee
The zero mode integral is 
\be
\int_0^{\b \sqrt{\b'}} dq_0 = \b \sqrt{\b'}\,.
\ee
We can now move onto the ghost contribution.  We have to exclude the $c$ zero mode when performing the path integral. This is because we do not gauge fix diffeomorphisms $\delta\tau = k$ for some constant $k$ (i.e. constant shifts). This means we are excluding a zero mode for $\delta\tau$, whose corresponding Grassmann field arising from the Faddeev-Popov procedure is $c$, so we should exclude the zero mode for $c$ as well.\footnote{We could also choose to gauge fix shifts by inserting a term like $\delta(\z(0))$, which fixes the origin of the circle. This is similar to fixing the residual freedom via the insertion of a vertex operator, although here we want to compute the vacuum amplitude with no insertions. The Faddeev-Popov procedure for this delta function would then produce an insertion of $c$ in the path integral which would soak up the $c$ zero mode. Thus we see that the number of $c$ ghost insertions agrees with the number of Killing vectors.} To evaluate the functional determinant we can expand the ghost fields in normalized, $\b'$-periodic eigenfunctions of $\partial_\t^2$, i.e. sines and cosines. $b$ has a zero mode $b_0/\sqrt{\b'}$, and we see that the insertion $4\b'^{-1}\int_0^{\b'} d\t\,b = 4 b_0/ \sqrt{\b'}$ picks it out. Thus the integral over the $b$ zero mode gives $4/\sqrt{\b'}$. The integral over the non-zero modes of $b$ and $c$ gives
\be
\int \mathcal{D}c \mathcal{D}b\,e^{-4\int_0^{\b'} d\t\, b\dot{c}} 
= \prod_{n=1}^{\infty} \f{8\pi n}{\b'} = \f{\sqrt{\b'}}{2}\,.  
\ee  
So altogether the ghosts contribute a factor of two, which is just a normalization chosen to cancel the $1/2$ that will come from dividing by time-reversal. The lack of $\b'$ dependence in the final answer can also be seen by scaling it away in the ghost action.

We are almost ready to assemble the pieces. The only thing left to do is divide out by the reparameterizations we did not fix. These were constant shifts in $\t$, which give a volume factor of $\b'$, and time reversal, which gives a factor of two. Putting everything together we get
\be\label{deformedPartitionFunction}
Z_{\l}(\b) = \frac{\b}{\sqrt{-8\pi \l}} \int_0^{\infty} \frac{d\b'}{\b'^{3/2}} \sum_{m\in \mathbb{Z}} \exp\left( \frac{1}{8\b'\l}(m\b - \b')^2 \right) Z(\b'),
\ee
where 
\be
Z(\b') = \int \mathcal{D}\Phi\, e^{-S[e=1,\Phi]},
\ee
is the undeformed partition function that depends on $\b'$ through the periodicity in $\t$. We see that the unit winding sector of \eqref{deformedPartitionFunction} gives the integral transform we found in \eqref{integraltransform} and hence the full deformed theory. 

An alternative route to a nonperturbative definition is to consider the worldline in static gauge  $\partial_{\t} X = 1$, instead of gauge fixing $e$. This is analogous to a proposal in two dimensions to view the $T\bar{T}$ deformation as a string worldsheet in static gauge \cite{McGough:2016lol, paperfuture2}. The deformed action then becomes
\be
S_E = -\frac{1}{8\l} \int_0^{\b'} e d\t \left(e^{-1} - 1\right)^2 + \int_0^{\b'} d\t e L_0(e,\Phi)\,.
\ee
In fact, this is similar to the deformed Schwarzian action \eqref{LSchdef}. To see this, write the action at temperature $\b'$ as 
\be
S_E = - \frac{1}{8\l} \int_0^{\b'} du\, \frac{\t'}{e^{\phi}}\left( \frac{e^{\phi}}{\t'} - 1 \right)^2 + \int_0^{\b'} du\, \frac{C}{2} \frac{\t'}{e^{\phi}}\left(\left(\frac{e^{\phi}\phi'}{\t'}\right)^2 - e^{2\phi} \right).
\ee
Now notice that if we covariantize the $\l = 0$ constraint $e^{\phi} = \t' \longrightarrow e^{\phi} = e^{-1} \t'$ and plug into this action we get
\be
S_E = -\frac{1}{8\l} \int_0^{\b'} du\, e\left(e^{-1} -1\right)^2 + \int_0^{\b'} du \,e \left[ \frac{C}{2}\left(\frac{(\partial_u (e^{-1}\partial_u \t))^2}{(\partial_u \t)^2} -  (e^{-1}\partial_{u}\t)^2 \right) \right],
\ee
where the term in square brackets is the covariantized undeformed Schwarzian action. (To covariantize one simply replaces $\partial_u\rightarrow e^{-1}\partial_u$ and adds a factor of $\sqrt{g}=e$.) While we have already shown in this section that any initial theory under our deformation can be understood as being coupled to a dynamical worldline, the argument here suggests that the deformed Schwarzian action can also be viewed, like the lower derivative theories, as coupling the Schwarzian action to a worldline in \emph{static} gauge.

\section{Discussion}
The Wilsonian paradigm applied to quantum mechanics implies a universal ultraviolet and a rich infrared, inverting the usual picture for quantum field theory. Deforming quantum-mechanical theories by operators built out of the stress tensor provides a calculable way to modify the ultraviolet and study the resulting physics. In holography, the richness of the infrared of quantum mechanics implies a huge landscape of asymptotically AdS$_2$ geometries. We have proposed deformations analogous to  $T\overline{T}$ in two dimensions which should isolate these exotic interiors and potentially provide a route to holography for more general spacetimes.

Many properties of theories deformed by our proposed flow \eqref{floweq} are exactly calculable, and we saw in section \ref{uvqm} that the deformed actions are rather universally found to be worldline actions. This suggests that our deformation couples the theory to an einbein, an interpretation which we made precise in section \ref{nonpert} by coupling the original theory to a theory of one-dimensional gravity.

For a particular sign of the deformation, the energy spectrum becomes complex beyond some $E_{n}$. It is often argued that one should truncate the spectrum beyond this point to maintain unitarity. In higher dimensions, this comes at the cost of spatial locality. In our case, there is no spatial locality to begin with, so this is a much less violent truncation. It defines a new quantum mechanics, where all observables of the theory can be expressed in the eigenbasis of the deformed Hamiltonian, $\mathcal{O} = \sum_{i=1}^n |E_i\rangle \langle E_i |$, where $E_n$ is the maximal energy. This is a principled yet somewhat impractical definition of the theory.

We now speculate on a few possible extensions to the work considered here.

\subsection*{\it Flowing from AdS$_2 \rightarrow$ AdS$_{d+1}$}
A thought that has surely crossed many minds is the following: if one sign of $T\overline{T}$ flow corresponds to flowing into AdS, shouldn't the other sign correspond to flowing out? Usually we think of AdS as being a self-contained, complete description -- and it is -- so this is a slightly strange question to ask. But in string theory the examples always arise from taking a near-horizon limit of branes in some asymptotically flat spacetime, so it is natural to ask how to recover the spacetime that we threw away in the near-horizon limit.

This idea was first seriously pursued in \cite{Giveon:2017nie} (see also \cite{Giveon:2017myj, Asrat:2017tzd, Chakraborty:2018kpr}), where for a concrete realization of AdS$_3$/CFT$_2$, a single-trace version of the $T\overline{T}$ deformation with the sign appropriate to flowing out was proposed and studied. The brane construction is of intersecting fundamental strings and NS-5 branes (compactified on $T^4$) with near-horizon AdS$_3$ times a compact manifold. The deformation was proposed to flow partially out, i.e. to move away from the fundamental strings but to stay near the NS-5 branes. This was proposed to deform the CFT$_2$ into a two-dimensional vacuum of little string theory, the six-dimensional worldvolume theory of NS-5 branes. 

Here we note that AdS$_2$, which is the case studied in this paper, is the universal near-horizon geometry of near-extremal black holes. In particular, JT gravity gives a universal description of the semiclassical physics of the states near extremality. Can we $T\overline{T}$ flow in quantum mechanics to discover the spacetime in which our near-extremal states are embedded? This can now be made precise by considering near-extremal black hole states in asymptotically AdS$_{d+1}$ spacetime. From the higher-dimensional point of view, we can fix to a charge-$Q$ sector and perform the $T^2$ flow of \cite{Taylor:2018xcy, Hartman:2018tkw} to some critical radius $r_c$ below which only near-extremal states fit. Of course, this flow can be reversed to get back to the AdS$_{d+1}$ boundary.\footnote{This kind of two-part flow was first introduced in \cite{Gorbenko:2018oov} to study de Sitter spacetime, although there the second part of the flow was modified by the addition of a term to reflect the de Sitter physics.}  But imagine dimensionally reducing the second half of the flow. This is a flow that stands on its own, and can be interpreted as a deformation of JT gravity, by an operator similar to the ones studied in the previous sections. 

To obtain the deforming operator, one can either dimensionally reduce the $d$-dimensional operator as in section \ref{rewrite}, or perform the analysis directly in the dimensionally reduced theory, as in section \ref{general2d}. To see how the latter analysis would work in a class of examples, consider the dimensional reduction of Einstein-Maxwell theory in $d+1$ dimensions (maintaining spherical symmetry) to $d=2$ dimensions:
\be
S = \f{1}{16\pi G} \int d^2 x \sqrt{-g}\left(\Phi^2 R + \lambda\left(\nabla \Phi\right)^2-U(\Phi) - f(\Phi) F^2\right) + S_{\text{bdry}}.
\ee
The kinetic term for the dilaton can be eliminated by a Weyl rescaling of the metric. By solving the equations of motion of the Maxwell field, one gets a smaller set of equations that is equivalent to the ones coming from the action above with $f(\Phi) = 0$ (see \cite{Almheiri:2016fws} for a nice discussion of this and other issues relating JT gravity to near-extremal black holes in higher dimensions). To extract the deforming operator at leading order in $N$, we only need the classical equations, so in the end we can consider
\be
S = \f{1}{16\pi G} \int d^2 x \sqrt{-g}\left(\Phi^2 R -U(\Phi) \right) + S_{\text{bdry}}.
\ee
Redefining $\Phi\rightarrow \sqrt{\Phi}$ puts this theory in the class considered in section \ref{general2d}.\footnote{As usual, the boundary terms are up to to the user and affect the deformation. To maintain finite quantities on an asymptotic AdS$_{d+1}$ boundary, the simplest choice is to dimensionally reduce the usual counterterms.} The deforming operator depends on the potential $U(\Phi)$ as in \eqref{KSflow}, and the potential depends on the dimension $d$. Thus, by choosing a particular $U(\Phi)$, we can flow out to an AdS$_{d+1}$ boundary for arbitrary $d$. All this really accomplishes, however, is embedding the AdS$_2$  region into AdS$_{d+1}$, unless the near-horizon theory is somehow defined as including the complex energy states resulting from the $T^2$ flow in the AdS$_{d+1}$ theory. 

\subsection*{\it SYK}
It is natural to ask about the application of our deformation, and more general ones constructed out of the stress tensor, to the SYK model. One can imagine performing the deformation either before or after averaging over disorder. In some simple cases these deformations can change the bilocal nature of SYK to trilocal, quadrilocal, and higher $n$-local theories. In the spirit of the techniques exploited in this paper, many quantities are still exactly calculable in the deformed theory. It becomes an interesting question to map out the space of integrable deformations to the SYK model, their effects on the Schwarzian effective action, and their nearly AdS$_2$ bulk interpretation. This is work in progress. 

\subsection*{\it Matrix models}
The deformation considered in this paper is naturally applied to one-dimensional theories. Recently, the authors of \cite{Saad:2019lba} have proposed a random matrix model description of quantities that are naturally computed from a two-dimensional bulk point of view. For example, the Euclidean wormhole with two boundaries of lengths $\b_1$ and $\b_2$ and of cylinder topology is the leading (in $L$) contribution to the matrix integral $\int DH e^{-L \Tr V(H)}Z(\b_1)Z(\b_2)$, where the integral is over $L\times L$ matrices drawn from the distribution $V(H)$. This is a theory in zero dimensions, and the correspondence with the bulk is \emph{not} ordinary AdS/CFT. Phrasing the effects of a $T\overline{T}$ deformation in terms of such theories is an interesting avenue to explore. One way is to travel from the two-dimensional bulk, to the one-dimensional boundary theory, to the zero-dimensional matrix model. This is spiritually the route followed in \cite{Saad:2019lba}, where the density of states of the one-dimensional Schwarzian theory, which is equivalent to the two-dimensional bulk, is used to read off a zero-dimensional matrix description. As we have seen, the density of states of the Schwarzian theory is modified by our deformation to \eqref{schwdens}. More general $f(H)$ deformations will modify it in different ways, and it would be fascinating to explore the types of matrix models one can access with these deformations. Truncating the spectrum for  deformations with a poorly behaved ultraviolet will imply a density of states with both left \emph{and} right edge, which is different than the doubly scaled matrix models studied in \cite{Saad:2019lba}.

\subsection*{\it $D0$ branes}
The gravitational theories discussed so far in this paper are motivated by top-down constructions, but it is not clear that they should be dual to any ordinary, unitary theories of quantum mechanics. For example, JT gravity has instead been proposed to be dual to a random matrix theory, which is not a unitary quantum-mechanical theory. On the other hand, the SYK model without disorder averaging is an ordinary quantum-mechanical system, but it does not appear to describe a bulk with a local Einstein-gravity like limit.

This brings us to the worldvolume theory of a stack of $D0$ branes. This is a one-dimensional matrix quantum mechanics with maximal supersymmetry and action given by the compactification of ten-dimensional $\mathcal{N}=1$ super Yang-Mills:
\be
S = \f{1}{2g} \int dt \Tr\left(\dot{X}^i\dot{X}^i+\Psi^T \dot{\Psi}+[X^i, X^j]^2-\Psi^T\g_i [\Psi, X^i]\right), \quad i = 1, \dots, 9\,.
\ee
The $X_i$ are nine bosonic $N \times N$ Hermitian matrices and $\Psi$ is a sixteen component $SO(9)$ spinor, which is also an $N \times N$ Hermitian matrix. The bulk theory is ten-dimensional Type IIA string theory, which has a low energy supergravity description. The matrix quantum mechanics above is also purported to describe $M$ theory in eleven dimensions with fixed lightlike momentum $P_- = N/R$, where $R$ is the radius of compactification. To decompactify, one takes the double scaling limit $N \to \infty$, $R \to \infty$ with their ratio fixed. So this is a top-down model of holography which has all the features we would want: the field theory dual is one-dimensional, and the bulk has a local Einstein-(super)gravity like limit with black holes, where questions about quantum gravity can be sharply studied.

The bulk theory dual to the $D0$ brane quantum mechanics is related to the dilaton-gravity models studied in this paper, but the spacetime is not asymptotically AdS. Nevertheless, the rules of holography are similar to AdS/CFT. To extract the physics of a finite Dirichlet cutoff one can study the flow of the bulk on-shell action and process it into a flow equation for the boundary action.

In the BFSS description separation in space is represented as separation in the matrices, so excising space should correspond to excising blocks of matrices. Or one can more generally consider the $f(H)$ deformations discussed in the introduction. These deformations -- especially in the way they will affect the ten- or eleven-dimensional gravitational theory -- are an interesting arena of exploration. 

%

\section*{Acknowledgments}
It is a pleasure to thank Dionysios Anninos, Tarek Anous, Jan de Boer, Victor Gorbenko, Tom Hartman, Raghu Mahajan, Edward Mazenc, Eva Silverstein, Douglas Stanford, and Herman Verlinde for useful conversations. JK is supported by the Delta ITP consortium, a program of the Netherlands Organisation for Scientific Research (NWO) that is funded by the Dutch Ministry of Education, Culture and Science (OCW). AR is supported by the National Science Foundation under Grant No. NSF PHY-1748958, the Department of Energy under DE-SC0009987, and by the Simons Foundation through the It from Qubit Simons Collaboration on Quantum Fields, Gravity and Information. ES is supported in part by NSF grant no. PHY-1316748 and Simons Foundation grant 488643. DG is supported by NSF grant 1125915.

\appendix
\section{$T^2$ deformations}\label{tangent}
The simplest deformations of quantum mechanics, i.e. $(0+1)$-dimensional QFT, are by powers of the stress tensor. The only possible quadratic contraction of the stress tensor is $T^2$.  Solving for the deformed energy spectrum for a $T^2$ flow is trivial:
\be
\f{\partial S}{\partial \lambda} =\int d\t \,T^2 \implies \f{\partial E}{\partial \lambda} = E^2 \implies E(\l) = \f{1}{E_0^{-1}-\l}\,,
\ee
where $E_0$ represents the original undeformed energy and we used the fact that $\langle T \rangle = E$. 
 It is interesting to determine the deformed Lagrangian. For 
 a free particle Euclidean Lagrangian $L_E = \dot{q}^2$, this can be done in two different ways. The first way is to write down a differential equation for the deformed Lagrangian $L_E(\lambda, q,\dot{q})$,  using $T=L_E - \f{\partial L_E}{\partial \dot{q}}\dot{q}$:
\be
\f{\partial L_E}{\partial \l} = \left(-\f{\partial L_E}{\partial \dot{q}}\, \dot{q}+L_E\right)^2.
\ee
Expanding the Lagrangian as 
\be\label{pertexp}
L_E(\lambda, q, \dot{q}) = \sum_{i=0}\l^i L_E^{(i)}
\ee
reduces the flow equation to 
\be
L_E^{(m+1)} = \f{1}{m+1}\sum_{i=0}^m\left(p^{(m-i)}p^{(i)} \dot{q}^2 - 2p^{(m-i)}\dot{q} L_E^{(i)} + L_E^{(m-i)}L_E^{(i)}\right)
\ee
for $p^{(i)} = \partial L_E^{(i)}/\partial \dot{q}$. For an initial free particle Lagrangian $L_E^{(0)} = \dot{q}^2$ one finds
\be
L_E^{(n)} = 2\,\f{(4n+1)!}{(n+1)!(3n+2)!} \,\dot{q}^{2n+2}\,,
\ee
which when inserted into \eqref{pertexp} gives
\be\label{3f2}
L_E(\l, \dot{q}) = \f{3}{4\l}\left(-1+_3\hspace{-1mm}F_2\left[-\f{1}{2},-\f{1}{4},\f{1}{4};\,\f{1}{3},\f{2}{3};\,\f{256 \l \dot{q}^2}{27}\right]\right).
\ee
As a functional of the original undeformed Lagrangian, this expression is precisely the same as $T\bar{T}$-deformed 2d Yang-Mills \cite{Conti:2018jho}. The connection exists due to the latter theory being quasi-topological, in particular the only non-vanishing component of $F_{\m\n}$ is $F_{01}$, so the deformation by $T\bar{T}$ is the same as deforming by $T^2$.

What about the more general case of interacting theories
? As shown above, the deformed energy spectrum is trivially calculable. In this sense the model is solved. But what if we wanted the deformed Lagrangian, something we could stick into a path integral? The differential equation above 
is difficult to solve directly. But we can transform the deformed Hamiltonian into a deformed (Lorentzian) Lagrangian, via
\be\label{transeqn}
L(\l, q, \dot{q}) = p(\l, q,\dot{q}) \dot{q} - H(\l, p(q, \dot{q}), q)\,.
\ee
This requires knowing the deformed Hamiltonian in terms of the canonical momenta. Since the eigenfunctions are unchanged under these flows, we have
\be\label{ansatz}
H(\l, p, q) =  \frac{1}{H_0(p, q)^{-1}-\l }\,.
\ee
We need to solve $dH(\l, p, q)/dp = \dot{q}$ for $p(\l, q, \dot{q})$ and plug into \eqref{transeqn}. 
This is soluble and gives an explicit Lagrangian, but its form is not illuminating. (For vanishing potential it reduces to \eqref{3f2} above.) This gives the second way of obtaining the deformed Lagrangian. 

For (0+1)-dimensional fermions, finding the new Lagrangian is much simpler. This follows from the fact that the kinetic term is topological and therefore does not contribute to the stress tensor. Suppose we start with the Euclidean Lagrangian of a complex fermion with standard kinetic term $L_E = \overline{\psi}\dot{\psi}+V(\psi,\overline{\psi})$. This has a Hamiltonian $H_0 = V(\psi, \overline{\psi})$. We deform this theory by a $T^2$ flow and Legendre transform \eqref{ansatz} to find the Lagrangian
\be
L_E = \overline{\psi}\dot{\psi} + \frac{1}{V^{-1}-\l }\,. 
\ee
\section{Deformation by conserved charges in $d=2$}\label{ccharges}
This appendix is about two-dimensional theories. It is known that deforming the conformal field theory of $N$ free bosons by $T\overline{T}$ leads to a Nambu-Goto action in $N+2$ target space dimensions \cite{Cavaglia:2016oda}. Adding interactions seems to ruin the Nambu-Goto form of the UV action \cite{Bonelli:2018kik}. We can instead consider deforming interacting bosons as discussed in section \ref{rewrite}, where we use the trace relation $T^\m_\m \propto \l T\overline{T}$ to substitute out for $T_{\p\p}$.  While this is the same deformation as $T\overline{T}$ for conformal theories, they differ for non-conformal theories. We will also pass from the local stress tensor to its nonlocal integrals. The reason for this is so that we can eventually factorize the deforming operator in momentum-energy eigenstates. Using the shorthand $E = \int T^\t_\t$ and $J=-i\int T_{\t\p}=-i \int T^{\t\p}$, the proposed flow is 
\be
\f{\partial S_E}{\partial \l} = \int d^2 x  \left(\f{E^2-J^2}{1/2-2\l E}\right).
\ee
We can consider this flow for self-interacting bosons, where we will solve for the deformed Lagrangian by transforming the Hamiltonian:
\be
S_E = \int d^2 x\sqrt{g}\left(\f{1}{2}\partial_\m\phi^i\partial^\m\p^i + V(\p_i)\right) \\
\implies T_{\m\n} = -\f{2}{\sqrt{g}}\f{\delta S_E}{\delta g^{\m\n}} = -\partial_\m\p^i\partial_\n\p^i + \f{1}{2}\d_{\m\n}\partial_\a\p^i\partial^\a\p^i+\d_{\m\n}V(\p_i)
\ee
The interactions can break the $SO(N)$ symmetry of the scalars. Setting the length of the spatial circle $L=1$, we have for the free scalar theory $J =-i T_{01} = i\dot{\phi}^i \partial_x \phi^i = \pi^i \partial_x \phi^i$ and $H = T_{00} = \f{1}{2}((\pi^i)^2+(\partial_x\p^i)^2)+V(\p_i)$:
\be
H(\l) = \f{1}{4\l}\left(1-\sqrt{1-8\l\left(\f{1}{2}\pi_i^2+\f{1}{2}(\partial_x\p_i)^2+V(\p_i)\right)+16 \l^2(\pi_i\partial_x\p_i)^2}\right)\\
\implies L_E(\l) = \f{1}{4\l}\left(1-\sqrt{\f{1-4\l (\partial_x\p_i)^2-8\l V(\p_i)}{1-4\l(\partial_x\p_i)^2}}\sqrt{\det(\delta_{\m\n}-4\l \partial_\m \p_i\partial_\n\p_i)}\right)\label{ngmetric}
\ee
Notice that this can be written as the Nambu-Goto action (plus a constant) but with nontrivial metric
\be
L_E(\l) =  \f{1}{4\l}\left(1-\sqrt{\det (\partial_i X^\m \partial_j X_\m})\right)\hspace{-1mm},\,\,\, g_{\m\n}= \delta_{\m\n}\sqrt{\f{1-4\l (\partial_x\p_i)^2-8\l V(\p_i)}{1-4\l(\partial_x\p_i)^2}}\,.
\ee
To recover \eqref{ngmetric} we fix to static gauge $X^0 = \t$, $X^i = \phi^i$, $X^{N+1} = x$. Notice that for vanishing potential the metric becomes trivial. This type of deformation mimics the quantum-mechanical deformations considered in section \ref{uvqm}, where we always obtained a worldline action in the ultraviolet with a target space metric set by the potential. 

\small
\bibliographystyle{ourbst}
\bibliography{TTbarBib}

\end{document}